# On the limitations of CFD modelling of flow boiling at high flow velocities and high heat fluxes

*Boštjan Končar[1,2], Matej Tekavčič[1], Aljoša Gajšek[1,2], Martin Draksler[1], Joris Fellinger[3], Marianne Richou[4]

[1] Jožef Stefan Institute, Ljubljana, Slovenia

[2] Faculty of Mathematics and Physics, University of Ljubljana

[3] Max Planck Institute for Plasma Physics, Greifswald, Germany

[4] CEA Institute for Magnetic Fusion Research (IRFM), St Paul-lez-Durance Cedex, France

*Corresponding author:

E-mail address: bostjan.koncar@ijs.si

**Abstract**

The ability of computational fluid dynamics (CFD) models to predict flow boiling at high heat flux and high flow velocity conditions has been investigated. High heat fluxes of about 10 MW/m$^2$ and high flow velocities of about 10 m/s typically appear in water cooling channels of divertor target elements in fusion reactors. In particular, the heat flux partitioning model used in the two-fluid CFD formulation was studied. CFD simulations of flow boiling in realistic divertor target cooling channels were performed and compared with conservative single-phase simulations. The predictive capability of CFD models for boiling was evaluated using experimental data, covering a wide range of flow velocities and heat fluxes. Existing CFD models correctly predicted void fraction and wall temperature at low flow velocities, but showed physically irrelevant results at higher velocities (above 3 m/s) leading to wall temperature overestimation. The study identified the wall heat flux partitioning model as the main contributor to the mispredictions and thoroughly discussed the main modelling shortcomings. By analysing the impact of operating conditions, some key boiling parameters, and state-of-the-art heat flux partitioning models, improvements of the model parameters are proposed. The simulations and model analyses are performed within the framework of the ANSYS CFX code and the results are compared with flow boiling experiments in uniformly and top-heated flow channels.

**Keywords:**

**flow boiling, CFD, heat flux partitioning, high flow velocities, wall temperature, void fraction**

# 1 Introduction

Nucleate flow boiling is widely recognized as a highly efficient heat transfer mechanism in many engineering systems. Its applications are pivotal in water-cooled nuclear reactors and for the cooling of components in nuclear fusion reactors. Flow boiling is particularly efficient to evacuate high heat fluxes, but it is very challenging when operating close to the critical heat flux (CHF), which triggers a sudden deterioration of heat transfer that may lead to severe damage of the structures. Control and accurate prediction of flow boiling, specifically at high heat fluxes, is therefore essential to mitigate these challenges.

In experimental fusion reactors [1],[2],[3] efficient cooling of high heat load components, such as a divertor, is vital for their operation. The divertor target surfaces, which are exposed to extremely high heat fluxes from incident plasma particles (up to 10 MW/m$^2$ during normal operation and up to 20 MW/m² during transients), are actively cooled by water channels [4]. These channels operate in either single-phase (forced convection) or two-phase (flow boiling) heat transfer regimes, with the latter being several times more effective. Their heat removal capability can be significantly enhanced, either by increasing the coolant flow rate or by applying geometric heat transfer promoters in the channels [5]. Operation in boiling regime at high heat fluxes and high flow velocities with sufficient margin to CHF is challenging and has been the subject of extensive experimental and numerical research.

To remove extremely high heat fluxes from the divertor, tubes with twisted tape inserts or hypervapotron channels are used to increase the heat transfer. Experimental studies were carried out to develop heat transfer correlations for flow boiling and CHF in hypervapotrons [6],[7]. Heat transfer performance of various cooling concepts for the ITER divertor, including twisted tapes and tubes with helical fins were investigated by Smid et al. [5]. Boscary et al. [8],[9] performed CHF experiments under one-side heating conditions in different water-cooled channels, i.e. smooth tubes, screw tubes, tubes with twisted tapes, and hypervapotrons. Based on dimensionless analysis, they have proposed dedicated CHF correlations for one-side heating conditions. An extensive set of experimental data with corresponding pressure drop and heat transfer correlations for horizontal flow boiling is provided in [10],[11]. More recently, Yan et al. [12] and Zhu et al. [13] investigated subcooled flow boiling in tubes with twisted tapes under high heat fluxes. They performed experiments and assessed a wide set of heat transfer correlations, particularly at high mass flow rates and high inlet subcooling relevant to divertor conditions. Despite numerous experimental and theoretical studies, the description of flow boiling lacks generality in terms of capturing the fundamental boiling physics. The design of components therefore still relies on costly experiments and empirical correlations valid only under specific operating conditions and geometries. This applies in particular to the flow boiling at high heat fluxes and high mass flow rates that are the focus of this study.

Recent advances in computational fluid dynamics (CFD) enable numerical simulations of boiling phenomena on a more fundamental level. The two-fluid modelling approach is widely used in nuclear community for simulations of boiling in realistic, mostly vertical flow channels [14],[15],[16],[17],[18],[19]. The state-of-the-art models in CFD codes are not completely suited for simulations of divertor cooling conditions in fusion applications, which are characterised by much higher heat fluxes and flow velocities. Flow boiling simulations in these conditions were performed by adjusting the modelling parameters [20],[21],[22],[25],[26] or by using fit-for purpose semi-mechanistic models [23],[24]. Modelling of flow boiling with the two-fluid framework is rather complex, as it has to capture the two-phase flow phenomena in the bulk flow (bubble interactions and condensation in the subcooled liquid) and the nucleate boiling on the wall-to fluid interface. The latter is modelled by a heat flux partitioning model, which divides the total heat flux conducted by the channel wall into a vapor generation part and a liquid heating part. Correct modelling of the partition between the vapor generation and subcooled liquid heating is essential for predicting the wall temperature. Heat flux partitioning models can be mechanistic or semi-mechanistic. Semi-mechanistic models are more engineering-oriented and are based on the augmentation of the overall heat transfer coefficient due to boiling [23]. On the other hand, mechanistic models try to capture the basic bubble nucleation phenomena, such as bubble departure diameter, departure frequency and density of active nucleation sites [27],[28].

As mentioned, CFD models do not yet reach a sufficient level of generality. Their use outside the range of experimental validation can yield unreliable results. Consequently, the inability to accurately predict boiling phenomena can lead to over-conservative engineering solutions, such as designing cooling channels for single-phase convective heat transfer [29]. The main objective of this study is therefore to evaluate the capability of existing models to simulate boiling at high heat fluxes and mass flows, typical for the cooling of divertor targets in fusion reactors [4]. In particular, the focus is on the mechanistic heat flux partitioning model, which dictate the prediction of wall temperature. A significant overestimation of the wall temperature during boiling at high flow velocities (above 3 m/s) was observed with existing heat partitioning models [27], widely used in CFD codes. This study systematically analyses the validity of the assumptions and the behaviour of model parameters under wide range of operating conditions. By analysing the state-of-the-art heat flux distribution model [30], possible improvements of the model parameters are proposed. The simulations and model analyses are performed within the framework of the ANSYS CFX code [31] and the results are compared with flow boiling experiments in uniformly and top-heated flow channels.

## 2 Flow boiling model

The Eulerian two-fluid modelling approach is used to simulate the flow boiling. The two-fluid model assumes that both phases share the same pressure. Momentum, mass and energy conservation equations are solved separately for the liquid and gas phase [19]. The liquid phase is modelled as a continuous phase and the vapour bubbles are modelled as a dispersed phase. Turbulence effects in the liquid phase are modelled by the k-ω SST model [32], while no turbulence is modelled in the dispersed vapour phase - a zero-equation model is applied [31]. The effects of the vapour bubbles on the liquid turbulence are modelled by the Sato approach [33].

Interactions between the liquid and vapour phases are modelled by additional interfacial transfer terms in conservation equations. Heat transfer between the two phases in the bulk flow was modelled by the Ranz and Marshall correlation [34]. Momentum transfer between the phases included only the two main interfacial forces: drag force modelled by the Schiller-Nauman model [35] and turbulence dispersion force using the Favre Averaged Drag model [36]. Bubble lift force and wall lubrication force models were not considered in these simulations as they do not significantly affect the heat transfer on the wall (on the solid-fluid interface). The two-fluid simulations were carried out with the ANSYS CFX, code version R2 2021 [31]. In this study, the main focus is given on the modelling of heat and mass transfer on wall that are modelled by the heat flux partitioning model.

### 2.1 Mechanistic heat flux partitioning model

Subcooled flow boiling regime up to the saturation is considered in the present work. Subcooled boiling describes the condition where the wall temperature is sufficiently high to trigger boiling on the heated wall surface, while the surrounding liquid is subcooled. The heat is transferred directly from the wall to the subcooled liquid, partly to heat up the liquid phase and partly to generate vapour bubbles. Moreover, condensation of the saturated vapour bubbles in the subcooled liquid in the core flow also increases the liquid temperature. In the state-of-the art CFD codes, these boiling conditions are often considered with the so-called mechanistic model for heat and mass transfer. Mechanistic approach means that the basic physical processes and parameters involved in boiling at the heated wall are incorporated into the model, i.e. bubble formation on the wall, bubble size, bubble departure frequency, nucleation site density, etc. The heat flux transferred from the heated surface is partitioned into the three contributions, as follows from the standard RPI (Rensselaer Polytechnic Institute) model [27]

$$q_w = q_{cl} + q_q + q_e, \qquad (1)$$

where $q_w$ is the total wall heat flux, $q_{cl}$ is the convective heat flux to the subcooled liquid, $q_q$ is the quenching heat flux and $q_e$ is the evaporative heat flux. Figure 1 shows an overview of the model and its main parameters in ANSYS CFX [31].

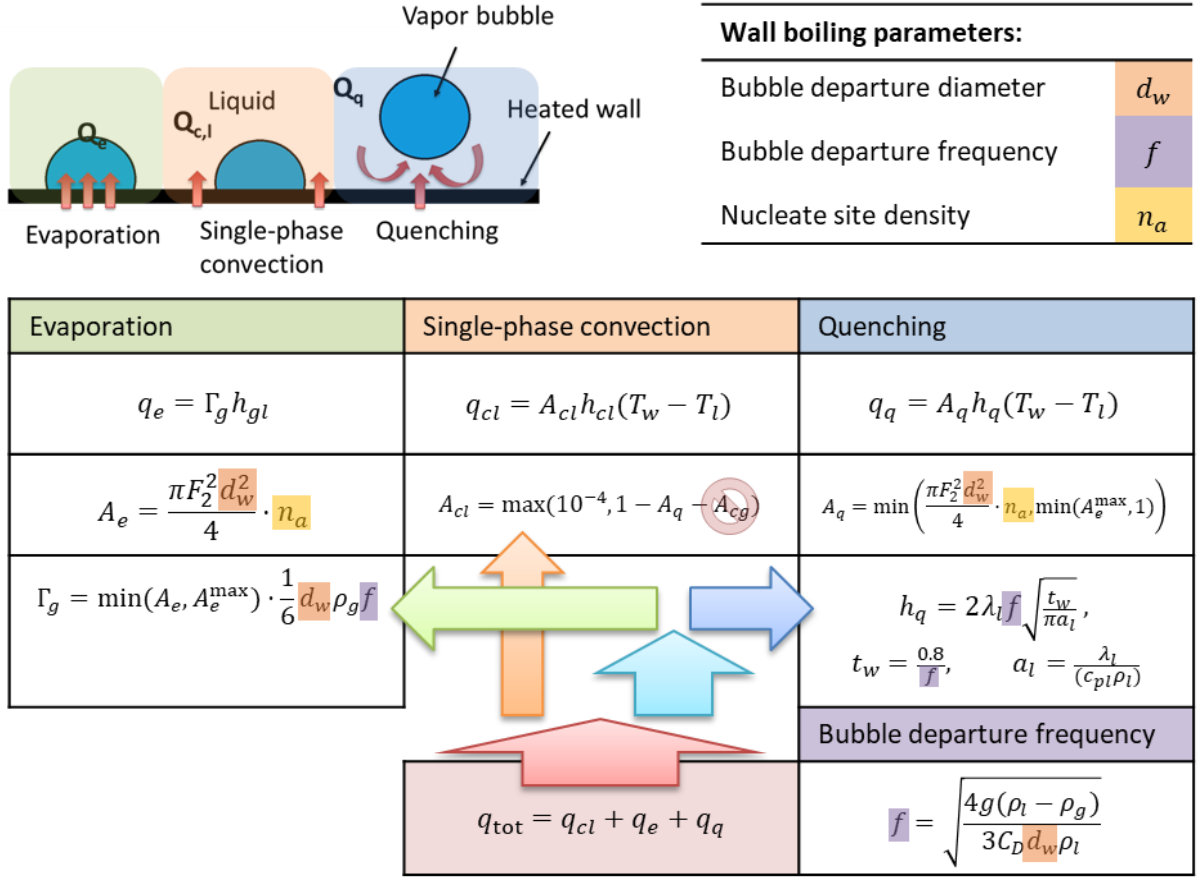

Figure 1: Overview of wall heat flux partitioning model in ANSYS CFX [31].

In the model, the heated wall surface is divided into two regions: one influenced by nucleating vapour bubbles ($A_q$), and the remaining, undisturbed part of the wall, covered by liquid, where heat transfer occurs solely through convection ($A_{cl} = 1 - A_q$). These area fractions, appearing in the subsequent equations (2) to (4), are given in dimensionless form, so they represent a fraction of the total heated wall surface. Assuming boiling conditions are not close to CHF, the part with convective heat transfer to vapour can be neglected ($A_{cg} \to 0$). In ANSYS CFX [31], the evaporation heat flux is expressed using the area fraction $A_e$, which is essentially equal to the area fraction $A_q$, but it allows for a different upper limit value $A_e^{\max}$. The modelling of these influence area fractions and their parameters is explained in more detail in Section 5.

The single-phase liquid convective heat flux is expressed as

$$q_{cl} = A_{cl} h_{cl} (T_w - T_l), \qquad (2)$$

where $h_{cl}$ is the single-phase heat transfer coefficient between the liquid and the wall at the surface fraction $A_{cl}$, and $T_w$ and $T_l$ are the temperatures of the wall surface and the near-wall liquid, respectively. The quenching heat flux $q_q$ represents the heating of the colder bulk liquid that fills the empty space left after the bubbles detach from the wall

$$q_q = A_q h_q (T_w - T_l). \tag{3}$$

Here $h_q$ represents the quenching heat transfer coefficient, which depends on the bubble departure diameter $d_w$ and the bubble departure frequency $f$. The quenching heat transfer model is discussed in sections 5.1 and 5.2. The evaporation heat flux is defined as

$$q_e = \Gamma_g h_{lg} = V_d\, n_a\, \rho_g\, f\, h_{lg} = A_e \cdot \frac{1}{6} d_w \rho_g\, f\, h_{lg}, \tag{4}$$

where $\Gamma_g$ is the evaporation mass flux, $V_d$ is the volume of the departing bubbles, $n_a$ is the active nucleation site density, $\rho_v$ is the vapour density and $h_{lg}$ is the latent heat of evaporation. Usually the bubbles are assumed to be spherical, so the volume $V_d$ of the bubble departing the wall can be written as $V_d = \pi d_w^3 / 6$. Here $d_w$ denotes the bubble departure diameter.

## 2.2 Basic boiling parameters

The heat flux partitioning model determines the three heat flux contributions based on the modelling of key boiling parameters. In this study a basic set of boiling parameters from the ANSYS CFX code is used [31]. The range of validity of the boiling parameters used in the CFD simulations is given where available.

**Bubble departure diameter**

Most commonly, the correlation of Tolubinsky-Kostanchuk [37] is used to model the bubble departure diameter $d_w$. This also the default model in ANSYS CFX [31]:

$$d_w = \min\left(d_{\text{ref}} \cdot \exp\left(-\frac{\Delta T_{\text{sub}}}{\Delta T_{\text{ref}}}\right), d_{\max}\right). \tag{5}$$

The parameters in the model are dimensional ($d_{\max}$= 1.4 mm, $d_{\text{ref}}$=0.6 mm, $\Delta T_{\text{ref}}$= 45 K) and are chosen to fit the pressurized water data. The model is clearly not universal, as it does not depend on the wall heat flux or pressure, and the departure diameter is only a function of a liquid subcooling. The parameters in the model have to be adjusted on a case-by-case basis. In this study the default values are adopted.

**Bubble departure frequency**

The basic model for the bubble departure frequency $f$ in ANSYS CFX is the Cole model [39]:

$$f = \sqrt{\frac{4g(\rho_l - \rho_g)}{3 C_D d_w \rho_l}}. \tag{6}$$

Study of departure frequency models performed by Yoo et al. [40] shows that Cole model performs well only at lower Reynolds numbers (Re < 3125), independently of wall heat flux and Jacob number. There are also other models available in the literature [41], [42], [30] that perform better in the higher Reynolds number range, but still lack the general applicability. The correlation of Situ et al. [41] is based on experiments performed in water at atmospheric pressure at the inlet velocity varied from 0.48 to 0.94 m/s and heat flux changed from 60 to 206 kW/m². The model of Basu et al. [42] is also based on experiments using water, but at higher heat flux data covering the following range of parameters: pressure from 1 to 3 bar, mass flux from 124 to 926 kg/sm², inlet subcooling from 0 to 52°C and heat flux from 25 kW/m² to 1,130 kW/m². The most recent model proposed by Kommajosyula [30] is based on the energy limit of the thermal boundary layer and should better capture the trends with increased flow conditions. It was validated using experimental databases of Basu et al. [42], Yoo et al. [43], and Richenderfer et al. [44], which cover a relatively narrow range of flow velocities up to 1 m/s, but the predicted bubble departure frequencies are far better fitted to the experimental data than in the case of Cole model.

**Quenching heat transfer**

The quenching heat transfer coefficient $h_q$ in eq. (3) is modelled according to Del Valle and Kenning [45]:

$$h_q = 2 f \sqrt{\frac{t_w \lambda_l c_{pl} \rho_l}{\pi}} = 2 \lambda_l f \sqrt{\frac{t_w}{\pi a_l}} \tag{7}$$

where $t_w$ is the bubble waiting time, $k_l$ is the thermal conductivity of the liquid, $c_{pl}$ is the specific heat capacity of the liquid, $\rho_l$ is the liquid density, and $a_l = \lambda_l/(\rho_l c_{pl})$ the heat diffusivity. The correlation was derived from the experiments at atmospheric pressure, inlet velocity of 1.7 m/s and wall heat flux ranging from 70 to 95% of CHF. The bubble waiting time $t_w$ is inversely proportional to the bubble departure frequency and is defined as $0.8/f$ in this study.

**Active nucleation site density**

The density of active nucleation sites $n_a$ in eq. (4) is modelled by the correlation of Lemmert-Chawla [38] and is related to the wall superheat by a rather simple relation

$$n_a = \left(m(T_w - T_{\text{sat}})\right)^p = n_{\text{ref}}\left((T_w - T_{\text{sat}})/\Delta T_{\text{ref}}\right)^p . \tag{8}$$

This model is implemented in ANSYS CFX using parameters $m = 185$, $p = 1.805$, and in the form shown on the right of (8) with $n_{\text{ref}} = 0.8 \cdot 9.992 \cdot 10^5$ [m$^{-2}$] and $\Delta T_{\text{ref}} = 10$ [K]. According to a recent review in [46], the correlation (8) is based on experiments with the coolant pressure between 0.1 and 0.2 MPa, but due to its simplicity and good performance, it is still one of the more widely used models in CFD simulations of subcooled boiling.

## 3 Simulation of flow boiling in a divertor target

As a typical example of flow boiling at high mass flow and high heat flux conditions, a cooling channel of a divertor target in a stellarator W7-X [3] has been analyzed. During normal operation, the top surface of the W7-X divertor target element is exposed to an incident heat flux of 10 MW/m². The heat flux is applied to the central part of the target, cooled by the water pipes, as shown in Figure 2. For boiling simulation, only the part of the target at the outlet section of the cooling pipe (marked by the red rectangle in Figure 2a) has been modelled. The highest wall temperatures are expected at the outlet part. Namely, single-phase simulations of the target element performed in our previous study [29] have shown that the local pipe wall temperatures on the outlet section of the pipe can be much higher than the local saturation temperature of the coolant. This implies that boiling can be expected in this part of the pipe.

A reduced geometry model for boiling simulations with a zoom-in of the numerical mesh is presented in Figure 2b. A diameter of the cooling pipe is 9 mm. The inlet temperature of 40ºC at the inflow to the reduced model is obtained from the prior single-phase simulation of the target element [29]. The inlet velocity and the operating pressure are set to 9 m/s and 1 MPa, respectively. As shown in Figure 2a, a constant heat flux of 10 MW/m² is applied in the middle of the model on plasma facing side (tiles colored in yellow). A heavy tungsten alloy (W3.5Ni1.5Cu) was used as the material for the tiles. The simulation model uses the basic set of boiling parameters as described in Section 2.

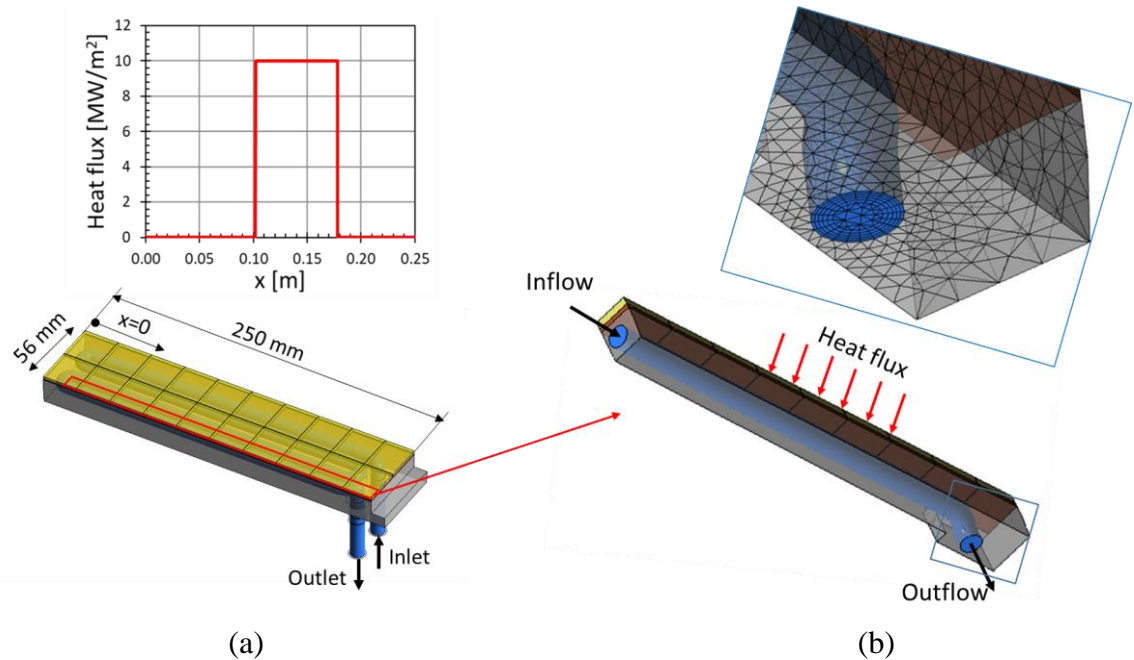

(a)                                                              (b)

Figure 2: (a) Target element of the W7-X divertor with the applied heat flux distribution. (b) Reduced geometry model of the outlet part of the cooling channel with a zoom-in of the mesh.

At the design stage, no wall temperature measurements are available, so the physical plausibility of the boiling simulations has been verified with the results of quasi single-phase simulations. The quasi single-phase simulation assumes that boiling does not occur at any time, even when the local liquid temperature exceeds the saturation temperature. In this case, the liquid continues to heat beyond the saturation temperature, like in the case of single-phase convection. In this respect, the single-phase simulation represents a conservative solution, i.e. the wall temperature calculated in the single-phase simulation should always be higher than the wall temperature calculated by the boiling model. This is due to the fact that boiling heat transfer is more efficient than the single-phase convection.

Figure 3 shows the simulated pipe wall temperature profiles at the solid-fluid interface, plotted along the axial line on the top of the pipe wall (red line in Figure 3 left). The single-phase simulation (SP) is represented by the red curve, while the blue dashed curve represents a two-phase (TP) simulation. The wall temperature predicted by the TP model significantly exceeds (by more than 100ºC) the SP simulation. In the boiling regime, this result can be considered as physically meaningless, as the wall temperature predicted by the two-phase model should be lower than the single-phase value.

The wall temperature in the two-phase regime could only be higher if the critical heat flux (CHF) would be exceeded. However, the existing boiling model does not include CHF modelling. Recently Baglietto et al. [47] proposed an approach to quantify the evolution of heat flux partitioning mechanisms towards CHF without its explicit modelling. However, the experimental tests from the literature [9] and Tong's empirical correlation [48] show that the CHF value for the flow boiling in the pipe under the considered conditions would be about 35 MW/m$^2$, which is far above the incident heat flux in the studied case.

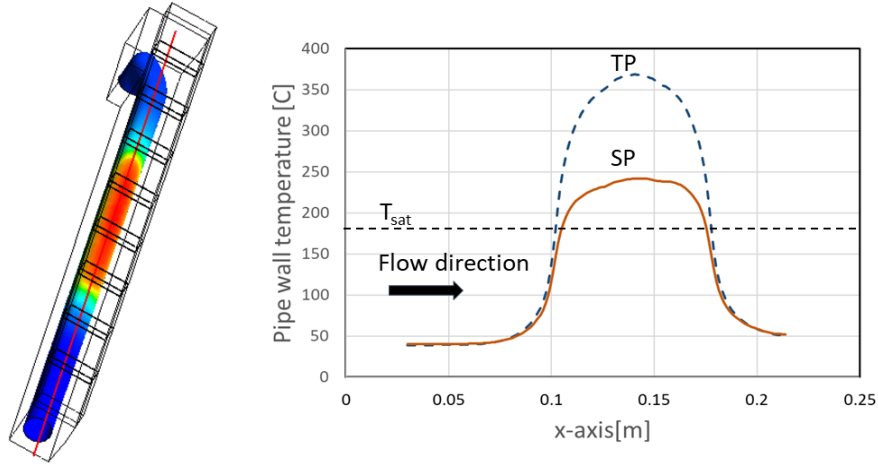

Figure 3: Single-phase (SP) and two-phase (TP) predictions of the pipe wall temperature distributions along the axial line on the top of the pipe wall (as shown on the left).

### 3.1 Analysis of a wider operating range

To investigate this issue, simulations of a wider range of heat fluxes and water inlet velocities were carried out. In this way, it can be determined at which conditions the model gives physically meaningful results. The results of SP and TP simulations were compared. Figure 4 shows the simulation results for different heat fluxes at a constant inlet velocity of 9 m/s. The red curves show the maximum temperatures of the tile top surface and the blue curves show the maximum pipe wall temperatures at different heat fluxes. The solid curves represent single-phase simulations ($T_t$ SP and $T_w$ SP thus denote maximum tile and pipe wall temperatures, calculated by the single-phase model) and the dashed curves represent two-phase simulations (with $T_t\_TP$ and $T_w$ TP denoting the maximum tile and pipe wall temperatures, respectively). The maximum void fraction at the wall is shown by the black circles. It can be observed that up to a heat flux of 7 MW/m$^2$, the maximum temperatures of the tiles and pipe wall are almost the same for SP and TP simulations, but then the temperatures calculated with the TP model start to exceed the SP values. This occurs as soon as the boiling is initiated, which is indicated by a small increase of the void fraction from 0.0007 to 0.004. Before this small jump in the void fraction, the flow in TP simulation is still in the single-phase convection regime, therefore both simulations predict the same temperatures. In the subcooled boiling region, i.e. at higher heat fluxes, the TP model should predict lower wall temperatures than the SP simulations. It is clear that at such high flow velocities and high heat fluxes, the TP model severely underpredicts the heat transfer between the heated wall and the boiling flow.

To assess the mesh effect, the TP simulations have been performed on two different numerical meshes, shown in Figure 4. The basic mesh, shown in Figure 2, is here denoted as the Coarse mesh and consists of 140,865 mesh cells. The fine mesh is refined in the near-wall region of the fluid domain and consists of 178,629 mesh cells. The purple and green curves in Figure 4 represent the temperature results on the fine mesh, with solid lines indicating SP simulations and the dashed lines TP simulations. On a finer mesh, the onset of boiling is shifted to higher heat fluxes, resulting in a slightly lower wall temperature and a higher void fraction (empty circles). However, this does not eliminate the problem, the resulting temperatures are still higher than in the SP simulation.

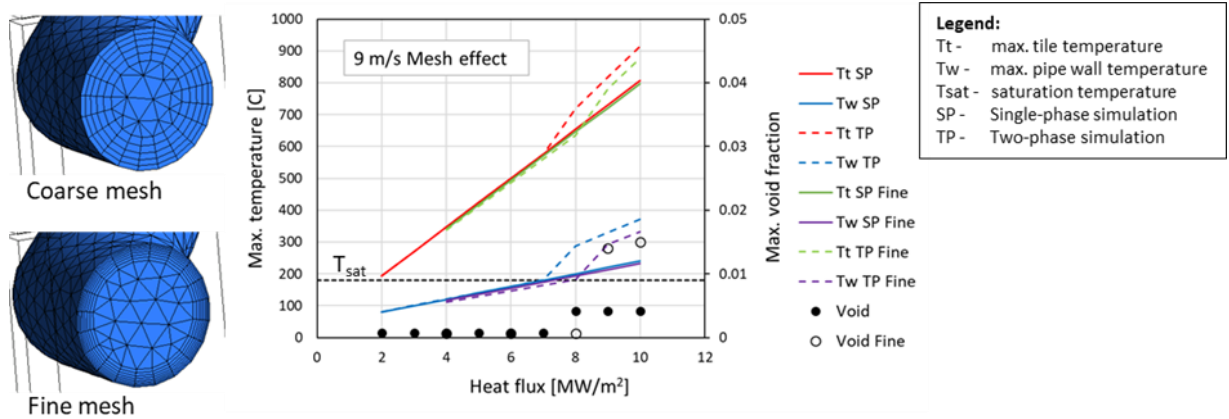

Figure 4: Maximum predicted temperatures and void fraction as a function of heat flux at an inlet velocity of 9 m/s. The effect of mesh is also shown.

The effect of the changed heat flux was also analyzed at lower inlet velocities of 1, 2, 3 and 5 m/s, as shown in Figure 5. All simulations are performed on the fine mesh. The simulations show that the TP model provides physically reasonable results for velocities below 2 m/s, where the maximum temperatures calculated by the TP model (dashed lines) are always lower than the SP values (solid lines). At higher velocities however, the TP model starts to exceed the temperatures predicted by the SP model. The turning point under these conditions is a velocity of 3 m/s. It is obvious that the existing TP model cannot properly simulate boiling at higher flow velocities.

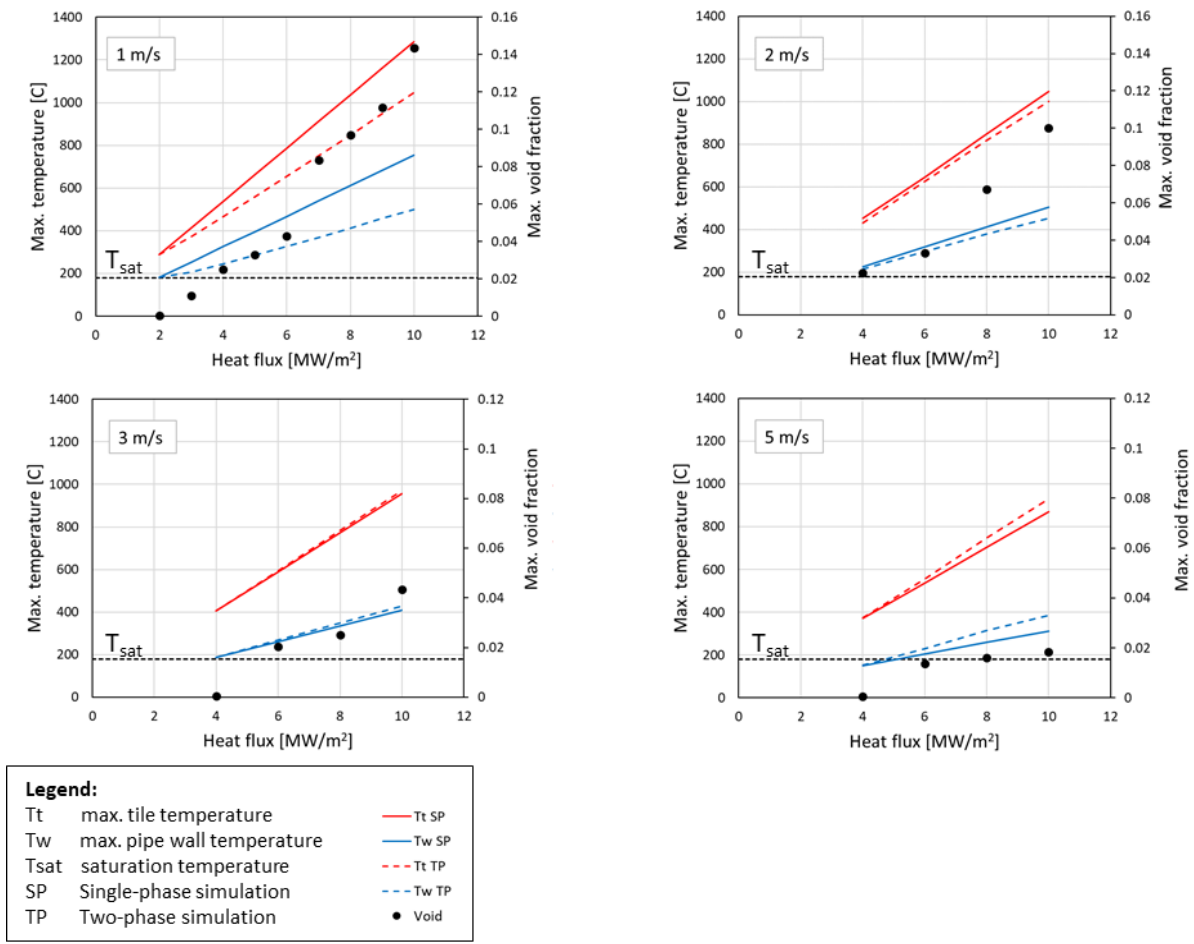

Figure 5: Maximum temperatures and void fraction as a function of heat flux at different inlet velocities of 1,2, 3 and 5 m/s

## 4 Validation on the available experimental data

Validation of the flow boiling simulations were performed on the Bartolomei experiments B1 [49] and B2 [50] in vertical pipe with uniform heating, and on the experiments of Komov et al. [11] in horizontal channel with top-side heating. The experiments that were the closest to the operating conditions or to the geometry of the cooling channels in fusion divertors were selected. To compare the geometry and operating conditions, the relevant dimensionless numbers are used in this study. The Reynolds number defines the level of flow turbulence as the ratio between inertial and viscous forces

$$Re = \frac{\rho_l u D_h}{\mu_l}, \tag{9}$$

where $D_h$, $u$ and $\mu_l$ are hydraulic diameter, flow velocity and dynamic viscosity of the liquid, respectively. The Weber number relates the inertia forces and the surface tension force

$$We = \frac{\rho_l u^2 D_h}{\sigma}, \tag{10}$$

where $\sigma$ is the surface tension. Following [52], $D_h$ is used as a characteristic dimension in We number. The Boiling number relates the wall heat flux $q_w$ with the mass flux $G$

$$Bo = \frac{q_w}{G h_{lg}}, \tag{11}$$

Typical operating parameters and dimensionless numbers for considered divertor conditions and experiments are given in Table 1. For comparison, the operating conditions in the reactor core of a typical pressurized water reactor (PWR) are also provided.

Table 1 Comparison of flow conditions and dimensionless numbers. Labels "B1" and "B2" represent two uniformly heated experiments of Bartolomei [49],[50].

| | | | Experiments | | |
|---|---|---|---|---|---|
| **Fluid** | Water (W7-X divertor) | Water (PWR) | Water (B1) [49] | Water (B2) [50] | Water (divertor-like) [11] |
| Pressure [MPa] | 1.0 | 15.7 | 1.5 | 4.5 | 1.0 |
| Sat. temp [°C] | 179.9 | 345.8 | 198.3 | 257.4 | 179.9 |
| Mean vel. [m/s] | 9.0 | 5.3 | 0.98 | 1.04 | 8.71 |
| G [kg/(m² s)] | 8965 | 4010 | 900 | 900 | 8700 |
| $q_w$ [MW/m²] | 10.0 | 0.5 | 0.78 | 0.57 | 2.6 to 16.7 |
| $q_w$ distribution | top-side heating | uniform | uniform | uniform | top-side heating |
| $T_{sub}$ [°C] (inlet) | 149.8 | 60.8 | 48.3 | 58.4 | 151.6 |
| $x_{th}$, (inlet) | -0.31 | -0.4 | -0.11 | -0.16 | -0.32 |
| $\rho_g/\rho_l$ (sat.) | 172.4 | 5.7 | 114.1 | 34.7 | 172.43 |
| $Re$ (inlet) | 101 200 | 424 000 | 118 117 | 101 874 | 53 957 |
| $Bo$ (sat.) | $5.5 \cdot 10^{-4}$ | $1.3 \cdot 10^{-4}$ | $4.4 \cdot 10^{-4}$ | $3.8 \cdot 10^{-4}$ | $1.6 \cdot 10^{-4}$ to $1.0 \cdot 10^{-3}$ |
| $We$ (sat.) | 15 319 | 37 142 | 526 | 537 | 7 337 |
| $Re/Re_{ref}$ | 1.00 | 4.19 | 1.17 | 1.01 | 0.53 |
| $Bo/Bo_{ref}$ | 1.00 | 0.24 | 0.80 | 0.68 | 0.29 to 1.72 |
| $We/We_{ref}$ | 1.00 | 2.42 | 0.03 | 0.04 | 0.48 |

In Table 1, the fusion divertor conditions in water cooling channels are taken as a reference case. Compared to the PWR conditions, the $Re$ numbers in divertor cooling channels and in the experiments are lower, but still highly turbulent. The $Bo$ number in divertor conditions is 4 times higher than in PWR, mainly due to a much higher heat flux. Therefore, the onset of boiling can be expected at the very beginning of the heated part of the divertor cooling channel, while the coolant flow in PWR core remains in a single-phase regime along the entire height of the PWR fuel rods during normal operation. Most dimensionless numbers in uniformly heated Bartolomei experiment B1 are similar, except for the *We* number, which is by two orders of magnitude lower than in the PWR or fusion divertor conditions. High *We* numbers are usually associated with highly turbulent flows dominated by inertial effects. The one-sided heated fusion experiments of Komov et al. [11] are the best approximation to the divertor conditions, as they also have a comparable *We* number.

### 4.1 Uniformly heated experiments

The experiments of Bartolomei [49],[50], although not fully matching the operating conditions of fusion divertor (heat flux 10 MW/m$^2$, flow velocity 9 m/s), have a similar geometry (pipe) and include the void fraction data and the measured wall temperature, which are both necessary for validation of the boiling model. The physical validity of the TP simulations was also verified by comparing the wall temperature results with the SP simulations.

The Bartolomei experiment B1 [49] was conducted in a uniformly heated vertical pipe with a diameter of 24 mm. The applied heat flux was 0.78 MW/m$^2$. The subcooled water entered at the bottom of the pipe at a velocity of 0.98 m/s. The operating pressure was 1.5 MPa. To simplify the simulation, the axial symmetry of the pipe is taken into account, hence a 2D simulation of half of the pipe's cross section was performed on a coarse 2D mesh with 20 by 200 mesh cells in radial and streamwise direction, respectively [51]. The basic boiling model, described in Section 2, was used for simulation of the experiment. To estimate the influence of modelling parameters, several parametric simulations were performed, where only one parameter was varied at a time while the others were kept constant. The effect of modelling parameters on the results is shown in Figure 6. The basic settings of the boiling model are denoted as "basic". The effect of different models for heat transfer in a bulk fluid has been analyzed. In the cases labelled "Nu=2" and "Nu=1500", the condensation Nusselt number was set to 2 and 1500, respectively in order to model very low or very high heat transfer in the bulk fluid. The "Hughmark" simulation indicates the case where the heat transfer is modelled by the Hughmark correlation [53] instead of the default Ranz-Marshall correlation [34] used in the "basic" model. In the "Db_2x" case, the bubble diameter in the bulk fluid is increased by a factor of 2 compared to the "basic" case. The amount of vapor generated on the wall depends on the size of the bubble departure diameter and on the number of nucleation sites. In the simulation "dbw_2x", the bubble departure diameter is increased by a factor of 2 and in the "Na_10x" simulation, the active nucleation site density $n_a$ is multiplied by a factor of 10 compared to the basic model. The results of average void fraction and the wall temperature along the pipe for different model variations are compared in Figure 6. Experimental values for the wall temperature are not available for this case, therefore the simulated temperatures are compared with the SP simulation results. Figure 6 shows that all models predict the average void fraction fairly similarly. The exceptions are the limiting cases with Nu=2 and Nu=1500. In the case of a high heat transfer coefficient (Nu=1500), the condensation of vapour bubbles is very intense, so the void fraction is lower, and more heat is transferred to the liquid. In the second case (Nu=2) the condensation heat transfer is very small, therefore the void fraction is higher. As can be observed, the predicted wall temperature is almost the same regardless of which variation of the model is used.

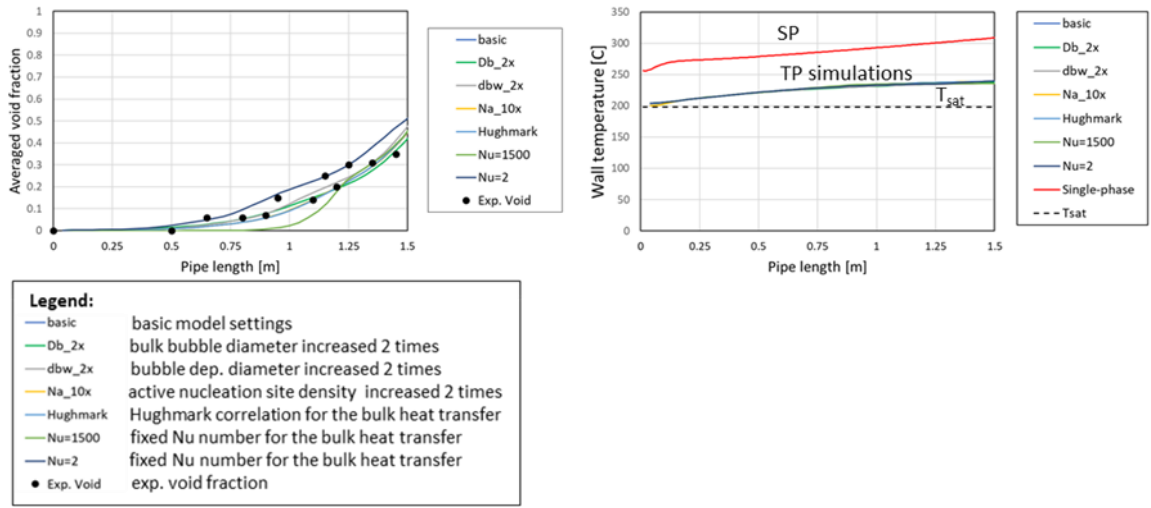

Figure 6: Experiment B1: Effect of modelling parameters on the averaged void fraction and pipe wall temperature.

In the experimental case B2 [48], in addition to the void fraction, the experimental data on the wall temperature and the average liquid temperature along the pipe are also available. The experiment B2 was carried out in a vertical pipe with a diameter of 15.4 mm, with a uniform heat flux of 0.57 MW/m$^2$. The operating pressure was 4.5 MPa, the mass flux 900 kg/sm$^2$ and the inlet subcooling 58°C. The basic model settings were used for the flow boiling simulations. The wall temperature, liquid temperature and void fraction predictions in Figure 7 show very good agreement with experiment. Similarly, good agreement has already been demonstrated also in a previous study by Krepper et al. [54]. Based on these results, it appears that the existing boiling model is suitable for the simulation of boiling flow for the tested conditions of pressure, temperature, velocity, tube geometry.

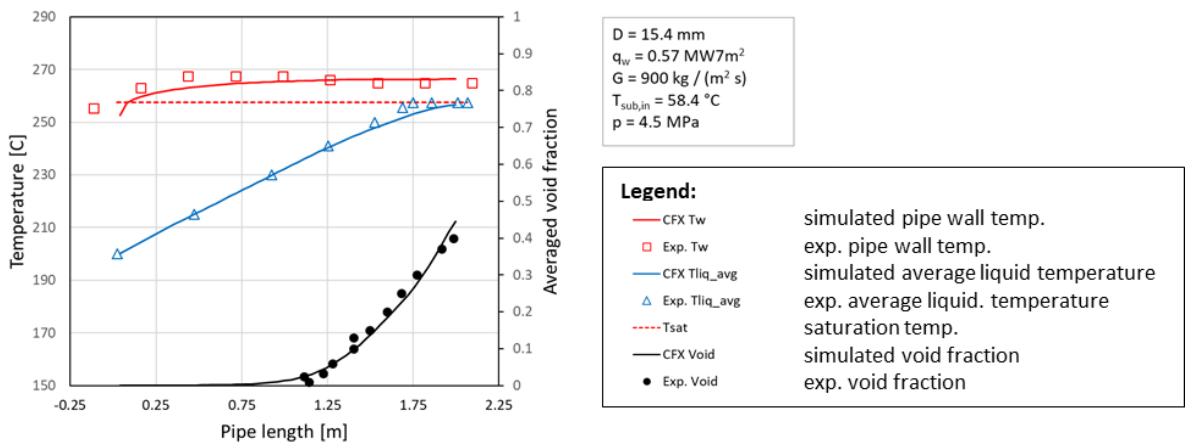

Figure 7: Experiment B2: Comparison of experimental and simulated wall temperature, averaged liquid temperature and void fraction.

### 4.2 Top-side heated divertor-like experiments

The experiments, similar in geometry and operating conditions to those in the divertor, were carried out by Komov et al. [11]. It should be noted that in these experiments only the temperatures in the monoblock and the pressure drop in the cooling tube were measured, while no data on the void fraction were available. The experimental test section under consideration, shown in Figure 8, consists of a top-side heated monoblock cooled by a 9 mm water tube with an inserted twisted tape (twisted tape coefficient k=0.9). Detailed description of the experimental setup is given in [11]. The conditions of the selected experiments are given in the last column of Table 1. The mass flux in the coolant tube was 8700 kg/sm$^2$ and the incident heat flux ranged from 2.6 to 16.7 MW/m$^2$. The SP and TP simulations are performed on the same mesh consisting of 3.06 million tetra elements with prism layers shown in Figure 8. The comparison of SP and TP simulations of the wall temperature with the experimental data at different heat fluxes is presented in Figure 9. The simulation "TP" denotes the results obtained with the base RPI heat partitioning model and basic set of boiling parameters, described in section 2.2. It can be seen that the SP and TP curves coincide as long as the wall temperature is below the saturation temperature and only single-phase convection is present. In the case of the flow boiling prediction, the TP simulation predicts higher wall temperatures than the SP simulation, but both are much higher than the experimental values. The TP model results can thus be considered as unphysical under these conditions.

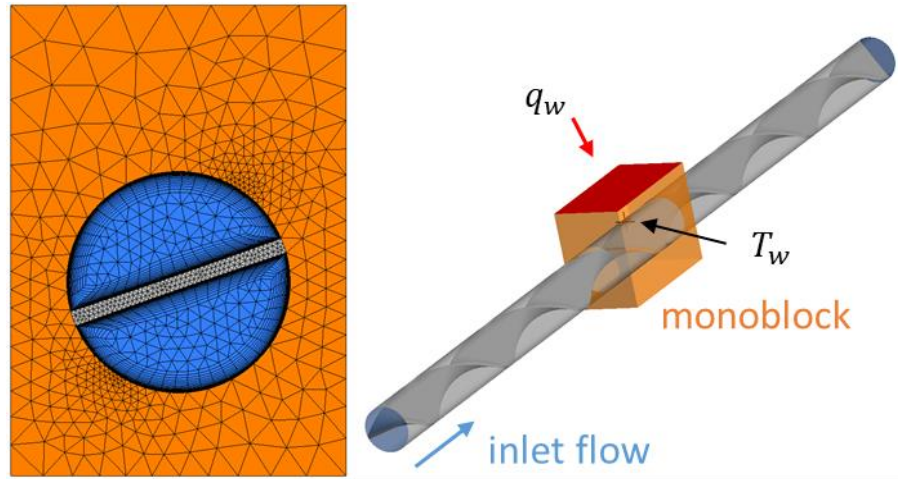

Figure 8: Geometry and front view of the mesh for the simulation of the experiment of Komov et al. [11] with top-side heated monoblock cooled by water tube with twisted tape insert.

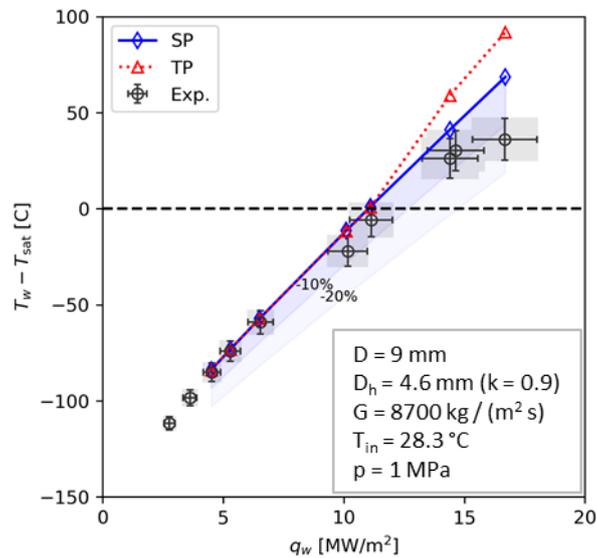

Figure 9: Comparison of experimental [11] and simulated wall temperatures for the one-side heating case at mass flux G = 8700 kg / (m$^2$ s) in pipe with twisted tape (k = 0.9).

# 5 Modelling issues in the RPI heat flux partitioning model

Several analyses were performed in an attempt to identify the reasons for erroneous predictions [51]. These, among others, included investigating possible user errors, issues with the turbulence modelling, modelling of the interface transfer in the bulk, and mesh accuracy near the wall. This study shows that the root cause for unphysical results at higher velocities stems from the wall heat flux partitioning model, which is applied outside the range of experimentally verified conditions and may incorrectly distribute the total wall heat flux between the two liquid heating contributions and the vapor generation part.

As discussed in more detail later on, the most problematic seems to be the model for the quenching heat flux. According to eq. (3), the quenching heat transfer coefficient depends on the bubble departure frequency $f$, which is by definition (see eq. (5)) based on the theory of buoyancy-drag force balance for pool boiling. Inevitably, this may lead to incorrect solution in conditions where the fluid velocity over the heated surface is high. The dependence on heat flux and flow velocity is not explicitly included in this model. It should be noted that $d_w$ and $f$ are interlinked (via the departure frequency model as in eq. (6)) and they appear in the models for evaporation and the quenching heat flux. The links between the individual sub-models within the overarching heat flux partitioning model are shown schematically in Figure 1.

The behaviour of the heat flux partitioning model at different conditions was investigated in detail on the case B1 (see Table 1). Due to the simpler geometry, this case is much easier to analyze due to uniform heating of the pipe wall. At the same operating pressure and inlet subcooling, the heat flux and flow velocity were modified in the simulations to reproduce the modelling issues observed in the divertor case:

- The wall heat flux was increased from 0.78 MW/m$^2$ to 3 MW/m$^2$.
- The flow velocity was increased from 0.98 m/s to 10 m/s.
- Note that the increase in flow velocity and heat flux results in significant increase of Re number (from 118 117 to 1 204 252) and We number (from 526 to 54 642)

For these analyses, a minimalistic mesh with 8x250 cells is used to facilitate fast simulation runs, as we are primarily interested in the heat flux partitioning model on the wall. Note also that for these modified B1 cases, the domain starts with an unheated pipe section at x = -0.5 m. Instead of an abrupt onset of heating at x = 0 m, a smooth heating function has been introduced (best shown in Figure 11d) to avoid numerical artefacts due to the sudden jump in heat flux and also to mimic the pipe wall heat flux distribution encountered in the divertor case (see Figure 3). The heating part of the pipe thus starts at approximately x=-0.1m.

The velocities and heat fluxes were varied with the goal to reproduce the erroneous predictions of wall temperatures described in the previous section. The SP and TP simulation results at varied heat fluxes for the low (0.98 m/s) and high (10 m/s) flow velocity are shown in Figure 10. The evolution of the wall temperature ($T_{\text{wall}}$), the liquid temperature in the first near wall cell ($T_{\text{fluid}}$), the saturation temperature ($T_{\text{sat}}$) and the near-wall void fraction ($\alpha_g$) along the pipe length are plotted.

For the two selected velocities, the wall heat flux ($q_w$) was varied in steps from approximately 1 MW/m$^2$ to 3 MW/m$^2$. The rationale for these is as follows. In the lower velocity case (0.98 m/s), already at 3 MW/m$^2$ a significant amount of water evaporates near the wall, as can be observed from the void fraction profile in Figure 10e, and increasing the $q_w$ would lead to CHF-like conditions that cannot be handled by the selected models. In the higher velocity case (10 m/s), at the lowest $q_w$ = 0.78 MW/m$^2$ no boiling occurs, as shown in Figure 10. Increasing the value to 2 MW/m$^2$ results in the onset of boiling shortly after about 1.2 m of the heated section of the pipe and immediately produces an erroneous result where the TP simulation of the wall temperature exceeds the SP simulation. A further increase to 3 MW/m$^2$ shifts the onset of boiling (and the error) to the beginning of the heated section, similarly to the case with a lower velocity of 1 m/s. As the error is already reproduced, no further increase of $q_w$ is necessary for this analysis.

The effect of increased heat flux on the wall temperatures at the low flow velocity (0.98 m/s) is shown in Figure 10a. In these cases, the predicted wall temperatures are below the SP results, suggesting that they are physically reasonable. However, taking a closer look at the local liquid temperature near the wall (see Figure 10c), the liquid temperature in the first near-wall cells exceeds the local saturation temperature at the low flow velocity. Given that the vapour temperature in the subcooled boiling regime is fixed at the saturation temperature, the liquid temperature should always remain below or at most equal to the saturation value these results are also questionable. Further analysis showed that modelling issues may already be present at lower flow velocities, but are masked by other effects. This point is discussed in more detail in the next section, where the flow velocity is varied at constant heat flux.

The effect of increased heat flux on the wall temperature at high flow velocity (10 m/s) is shown in Figure 10b. As mentioned above, increasing the flow velocity to 10 m/s at the low wall heat flux of 0.78 MW/m$^2$ is expected to result in a single-phase flow. Under these conditions, the convective heat transfer is able to cool the wall sufficiently and keep the wall temperature below the saturation temperature, which does not activate the wall boiling model. The modelling issues become evident at higher heat fluxes (2 and 3 MW/m$^2$). When the wall temperature increases above the saturation temperature and the wall boiling model is activated, the predicted wall temperatures exceed the single-phase results, which is physically impossible.

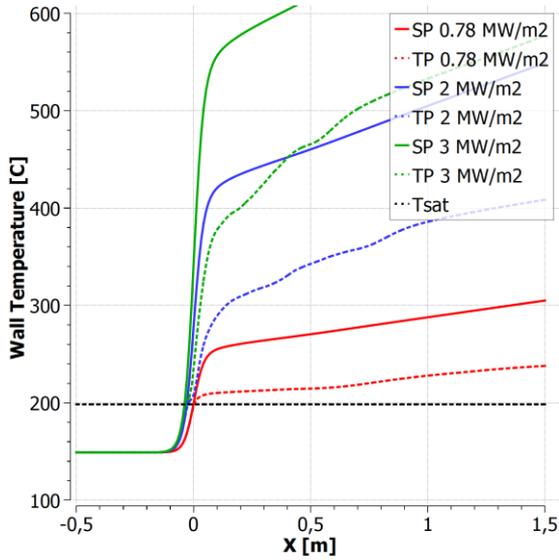
(a) $T_{wall}$ with varied $q_w$ at 0.98 m/s

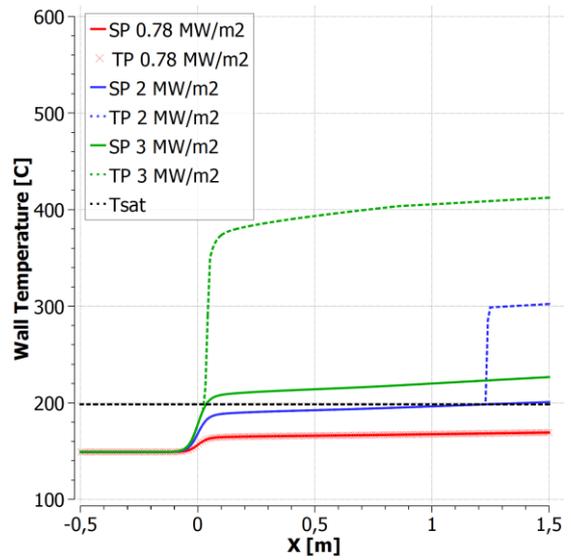
(b) $T_{wall}$ with varied $q_w$ at 10 m/s

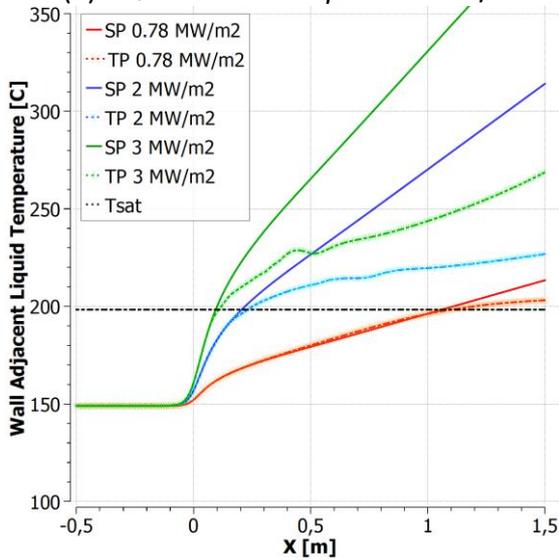
(c) $T_{fluid}$ with varied $q_w$ at 0.98 m/s

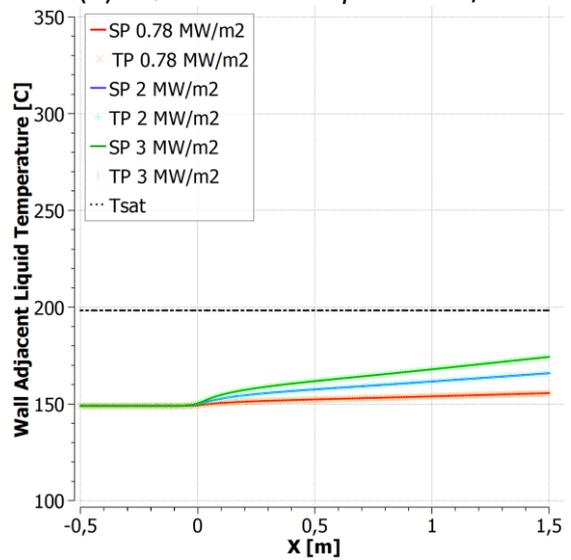
(d) $T_{fluid}$ with varied $q_w$ at 10 m/s

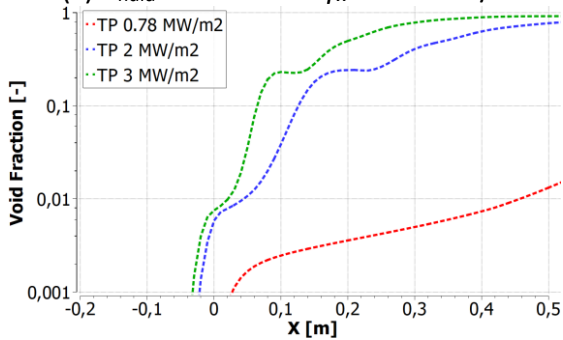
(e) $\alpha_g$ with varied $q_w$ at 0.98 m/s

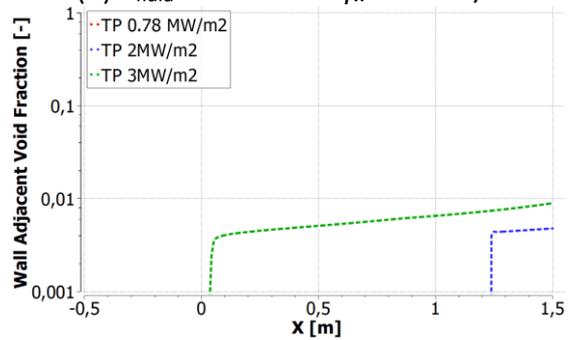
(f) $\alpha_g$ with varied $q_w$ at 10 m/s

Figure 10: Modified B1 case: Comparison of temperatures and void fractions between single-phase (SP) and two-phase simulations with boiling model (TP) at varied heat flux for the low flow velocity of 1 m/s (left column) and for the high flow velocity of 10 m/s (right column).

## 5.1 Effect of flow velocity on the heat partitioning model

The behaviour of the heat partitioning model at low (1 m/s) and high (10 m/s) flow velocities is studied at the constant heat flux 3MW/m². Wall and liquid temperatures, bubble influence area fraction $A_q$, heat transfer coefficients (HTC) and heat flux contributions are compared in Figure 11. Here, the heat transfer coefficients corresponding to the heat fluxes, i.e. SP convective $\text{HTC}_{cl}^{SP}$, and the TP convective $\text{HTC}_{cl}^{TP}$ and quenching $\text{HTC}_q^{TP}$, are defined as follows:

$$\text{HTC}_{cl}^{SP} = \frac{q_w}{(T_w - T_l)}, \qquad \text{HTC}_{cl,q}^{TP} = A_{cl,q} h_{cl,q} = \frac{q_{cl,q}}{(T_w - T_l)}. \tag{12}$$

Specifically, the two-phase HTC definitions include the corresponding influence area fractions for convective $A_{cl}$ and quenching part $A_q$, respectively, which enable better insight into the behaviour of the heat-flux partitioning model. For clarity, only the beginning of the heated length up to 0.5 m is shown and the results are compared with the single-phase simulation. The same labelling as in Figure 10 is used. Three main problems have been identified in the underlying heat flux partitioning method:

- The bubble influence area fraction $A_q$, where boiling and quenching occur, seems to be overpredicted as its value is being driven towards the upper limit of 1. This value implies that boiling occurs over the whole wall surface, i.e. resulting in only quenching and evaporation heat flux, without the convective heat flux contribution. Considering that this appears at the moment when the boiling model is activated, i.e. immediately after the wall temperature exceeds the saturation temperature (as shown in Figure 11a and b), this is difficult to justify theoretically.
- As shown in Figure 11c, the present model predicts rather similar values of the quenching heat transfer coefficient $h_q$ for the low flow velocity at 1 m/s (dashed blue line) and for the high flow velocity 10 m/s (full blue line). This is most likely incorrect, as it seems that the effect of inertia on the quenching heat transfer is not considered in the model.
- Underestimation of the evaporation heat flux $q_e$ itself, which is difficult to estimate under these conditions without the experimental data.

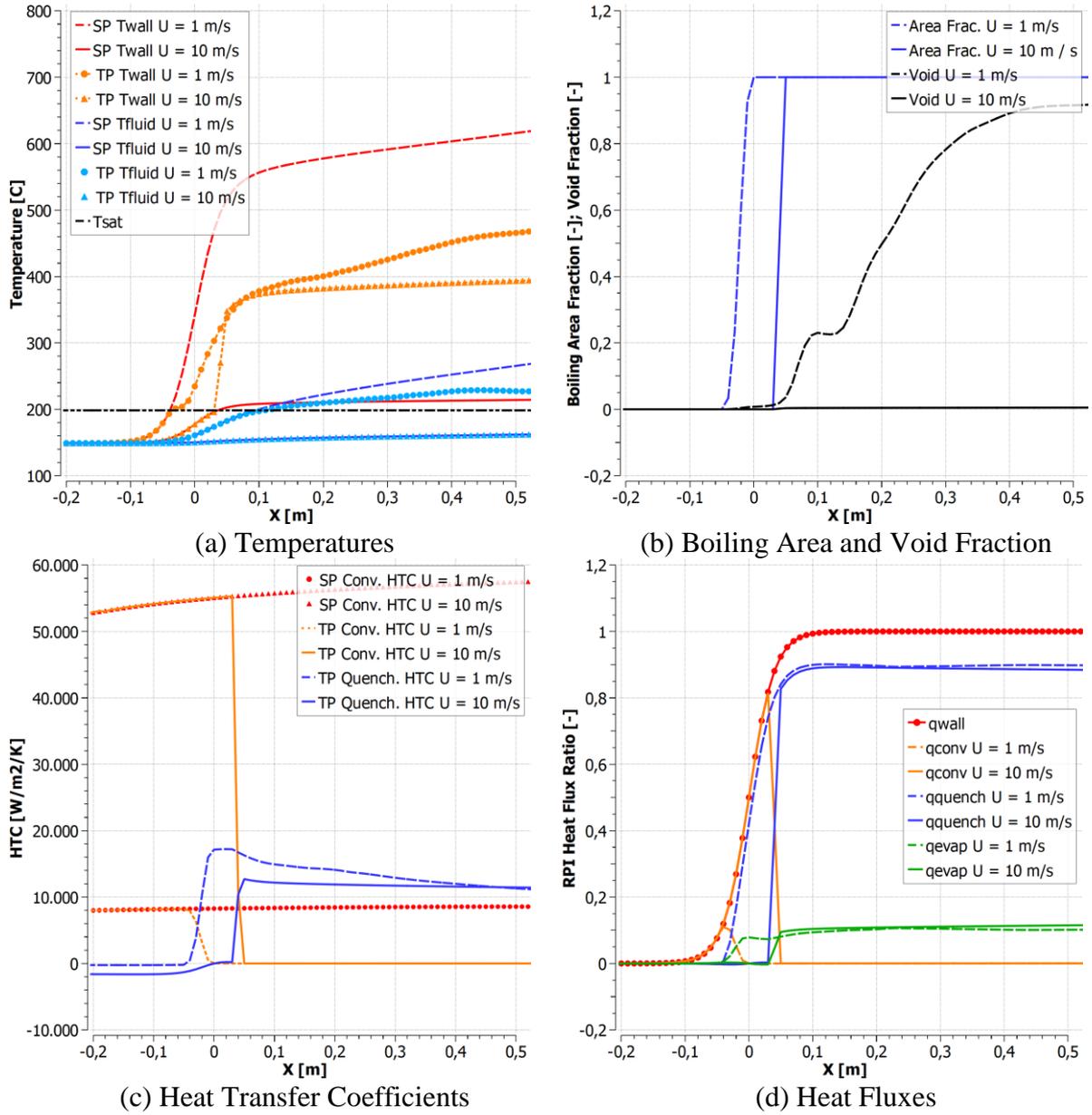

Figure 11: Modified Bartolomei B1 case: Behaviour of heat flux partitioning model at 1 m/s and 10 m/s at the same heat flux $q_w$= 3 MW/m². Comparison of wall ("Twall") and liquid temperatures ("Tfluid") (a) bubble influence area fraction $A_q$ (b), heat transfer coefficients (c) and heat flux contributions (d). Labels "qwall", "qquen", "qconv", "qevap" correspond to total wall, quenching, convective, and evaporation heat flux, respectively.

**Comments on the bubble influence area fraction $A_q$**

As shown by the overview of heat flux partitioning in Figure 1, the CFX models for the wall area fraction influenced by bubbles are essentially the same for the boiling $A_e$ and quenching $A_q$ fractions and differ only by the upper limit. The model for quenching fraction is defined as $A_q = \min\left[\frac{\pi F_2^2 d_w^2}{4} \cdot n_a, \min(A_{e,\max}, 1)\right]$ and is essentially a product of a single bubble influence area and the density of active nucleation sites $n_a$ on the heated wall. Here $\frac{\pi d_w^2}{4}$ represents the projected surface of the bubble at departure and $F_2$ represents the bubble influence area factor, by default set to 2. The value $A_q$ is limited either by the predetermined value $A_e^{\max}$ (with the default value of 0.5 in CFX), or by maximum fixed value of 1, which represents a wall totally covered by bubbles. Disregarding convection to vapour $A_{cg} \to 0$, the convective area fraction to liquid is defined as $A_{cl} = \max(10^{-4}, 1 - A_q)$, and is thus limited between the lower bound of $10^{-4}$ and upper bound of 1.

**Comments on the convective heat flux and convective heat transfer coefficient**

The wall heat transfer coefficient in the single-phase simulation (SP Conv. HTC), shown by the red dots (1 m/s) and squares (10 m/s) in Figure 11c, increases for about five times, as the flow velocity increases from 1 m/s to 10 m/s. Its values range between 50,000 and 60,000 W/m²K at 10 m/s and around 10,000 at 1 m/s and agree well with predictions of established HTC correlations for turbulent flows [55].

The prediction of the convective heat transfer coefficient $h_{cl}$ in the two-phase simulation (TP Conv. HTC; orange lines, dashed for 1 m/s and solid for 10 m/s) is comparable to the single-phase HTC curves until the boiling model is activated (after $T_{wall} > T_{sat}$). In the present case, the RPI model assumes that boiling region occupies the entire wall surface, immediately after the onset of boiling ($T_{wall} > T_{sat}$) and splits the heat flux between quenching (shown with blue lines) and evaporation heat flux only, while the convective heat flux instantly drops to zero.

Setting the prescribed upper limit for bubble influence to $A_{e,\max} = 1$, helped to reveal difficulty in modelling the quenching heat flux, as in this case the convective heat flux can be driven to the lower bound, and is effectively replaced by quenching heat flux, as shown Figure 11d. However, unlike the convective $h_{cl}$, the quenching $h_q$ is rather insensitive to the flow velocity in the model. Thus, a larger convective $h_{cl}$ at higher velocity 10 m/s is replaced by a much lower quenching $h_q$.

**Comments on the quenching heat flux and quenching heat transfer coefficient**

As shown in Figure 11c, very similar values for the quenching heat transfer coefficient $h_q$ are obtained for different flow velocities of 1 m/s and 10 m/s, respectively. This is in contradiction to the large difference in the convective heat transfer coefficient $h_{cl}$ (by a factor of 5) between the low and high flow velocity. In fact, the higher flow velocity case (10 m/s) results in an even slightly lower $h_q$. Furthermore, the $h_q$ values for both flow velocity cases (1 and 10 m/s) are comparable to the $h_{cl}$ value for the low flow velocity (1 m/s) case. Since the convective $h_{cl}$ is replaced by a comparable value of $h_q$ at 1 m/s, the overall heat-transfer capability remains the same between single-phase and boiling model, which hides the mentioned issues with wall temperature in simulation with boiling.

In the high flow velocity case (10 m/s), due to the dominant bubble influence area fraction $A_q$ and resulting negligible convective part $A_{cl}$, the combined heat transfer coefficient is dominated by the five times lower $h_q$. This significantly reduces the overall heat transfer and leads to erroneous over-prediction of the wall temperature that exceeds the SP simulation results.

The model for the quenching heat transfer in eq. (7) is rather insensitive to the change in the flow velocity. It depends on the bubble departure frequency $f$ and bubble waiting time $t_w$, which do not include an explicit dependency on either the flow velocity or heat flux. Namely, the existing Cole model [39] for $f$ is based on a time scale of a rising bubble with terminal velocity due to buoyancy in stagnant fluid (i.e. in pool boiling). The modelling of quenching heat transfer is found to be the most dominant source of erroneous wall temperature prediction, hence the influence of the main boiling parameters on the quenching is analyzed next.

## 5.2 Effect of boiling parameters

The quenching heat transfer and the partitioning of the heat flux is largely dictated by the three main boiling parameters in the heat partitioning model:

- Bubble departure diameter: $d_w$
- Bubble departure frequency: $f$
- Active nucleation site density: $n_a$

As shown in heat flux partitioning scheme in Figure 1, all three parameters are interrelated in the models for area fractions $A_e$ and $A_q$ ($d_w$ and $n_a$), evaporation heat flux $q_e$ ($d_w$, $f$ and indirectly in $A_e$) and quenching heat transfer coefficient $h_q$ ($f$, and indirectly $d_w$). Models for $d_w$ and $f$ are interdependent via the commonly used Cole model [39]. The complexity of the analysis is further enhanced by the fact that the $d_w$ model in eq. (5) and nucleate site density $n_a$ model in eq. (8) depend on the wall temperature, which is the output result of the RPI model (at a prescribed wall heat flux), resulting in interdependence that is resolved by an iterative process.

A sensitivity analysis of the heat partitioning model on the bubble departure diameter $d_w$ and frequency $f$ was performed for the modified B1 case with the heat flux of 3 MW/m$^2$ and the flow velocity of 10 m/s. To control the behavior of the heat-partitioning model, the values of $d_w$ and $f$ were set to artificial constant values, making them independent of the wall temperature. Note that the predefined values of $d_w$ and $f$ can sometimes be unrealistic and are merely used to demonstrate the effect of the quenching heat flux model on the overall results.

The influence of $d_w$ model is studied in Figure 12. In the figure, the reference value obtained by equation (5) is denoted by "dw ref", and the values reduced by a factor of 2 and 4 are denoted by "dw/2" and "dw/4", respectively. For the purpose of this analysis, the influence factor $F_2$ was set to 1. As shown in Figure 12a, the reduced bubble departure diameter results in a decreased wall temperature. A reduction of $d_w$ by 4 times also delays the onset of steep increase of T$_{wall}$, but the wall temperature still remains well above the values predicted by the single-phase model. Obviously, the smaller value of $d_w$ itself cannot resolve the unphysical prediction of T$_{wall}$. The delayed rise of the wall temperature most likely occurs due to the delayed jump of the bubble influence area fraction $A_q$ (smaller product of $d_w$ and $n_a$, which nevertheless still increases with T$_{wall}$) shown in Figure 12c. The effect of $d_w$ on the heat flux contributions is shown in Figure 12b. Ratios between the specific heat flux and total wall heat flux are presented. Reduced value of $d_w$ results in decreased convective $q_c$ (orange lines) and evaporation $q_e$ (green lines) heat fluxes and in an increased quenching $q_q$ heat flux (blue lines). Note that in the results shown here, the $A_q$ is limited to 0.9. When the limiting value of $A_q = 0.9$ is reached, the resulting RPI heat fluxes also reach their limiting values, which persist along the heated wall.

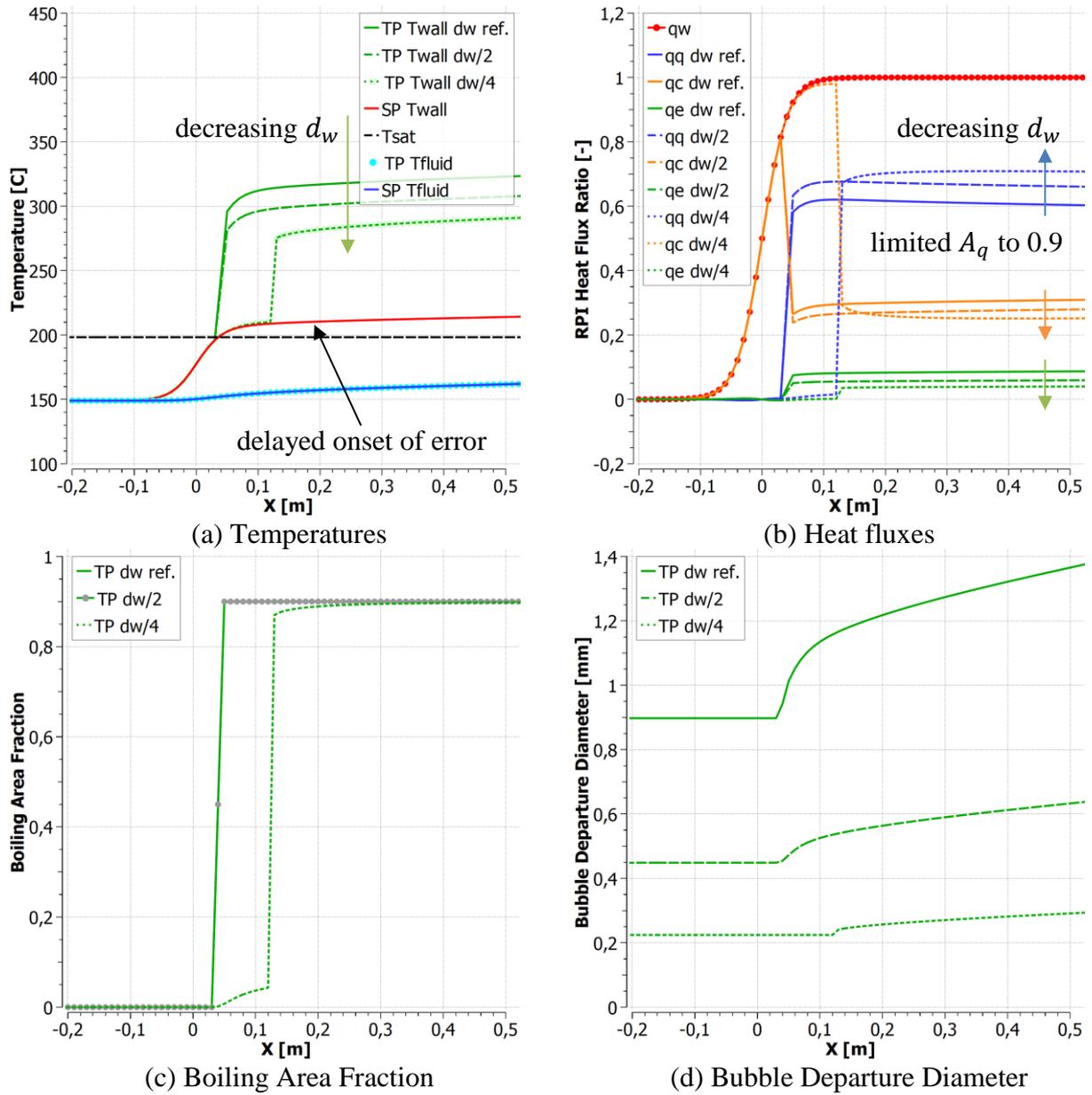

Figure 12: Modified Bartolomei B1 case with 10 m/s and 3 MW/m$^2$: Effect of decreasing bubble departure diameter $d_w$ on the wall temperature (a), heat partitioning (b), boiling area fraction (c), with the corresponding diameter values (d). Labels "qw", "qq", "qc", "qe" correspond to the total wall heat flux, quenching, convective, and evaporation heat flux, respectively.

In Figure 13, the influence of bubble departure frequency $f$ is analyzed. Here, the bubble departure diameter was fixed at a constant value of $d_w = 0.1$ mm, while different constant values of $f$ are imposed. Note that at a constant bubble diameter $d_w$ and constant fluid properties (e.g. gas density $\rho_g$ and liquid density $\rho_l$), the standard Cole model [39] in eq. (6) results in a constant value for the bubble departure frequency $f$ along the pipe wall. Furthermore, setting the $f$ independently of $d_w$, is similar to modifying the relationship between the two, in other words it mimics the use of an alternative model for the frequency. In Figure 13a, the default frequency value $f = 360\ s^{-1}$ (obtained from the Cole model using a constant $d_w$) was gradually increased (independently of $d_w$) by several orders of magnitude up to a maximum value of $f = 5 \cdot 10^5\ s^{-1}$. Note that these frequency values are used solely to investigate the model response and can be sometimes much higher than the measured data from the literature [57]. As can be observed in Figure 13a, the wall temperature $T_{\text{wall}}$ decreases with the increased bubble departure frequency $f$. Above a certain value of $f$ ($5 \cdot 10^3\ s^{-1}$), the resulting $T_{\text{wall}}$ drops below the single-phase (SP) prediction, hence correcting the unphysical behavior of the model.

RPI heat flux ratios normalized to the total wall heat flux are shown in Figure 13b. Instead of a sharp stepwise change in the axial profiles of the heat fluxes observed in Figure 12b, much smoother heat flux profiles can be observed at increased frequency of $5 \cdot 10^3\ s^{-1}$. Smoother heat flux curves result from the improved behaviour of $A_q$ and lower $T_{\text{wall}}$, which in combination with a constant value of $d_w = 0.1$ mm affects (decreases) the product of nucleation site density $n_a$ and bubble diameter $d_w$ in the calculation of $A_q$, as shown in Figure 13c and d, the $A_q$ is not driven towards its upper limit, which would flatten the axial profiles as observed before. Due to the smooth change of the bubble influence area fraction $A_q$, the convective heat flux does not drop drastically. The increased bubble departure frequency $f$ also enhances the quenching (Eq. 7) and the evaporation heat flux (Eq. 4). These modifications enhance the heat transfer during the subcooled boiling and reduce the wall temperature below the SP prediction, thus providing physically plausible results.

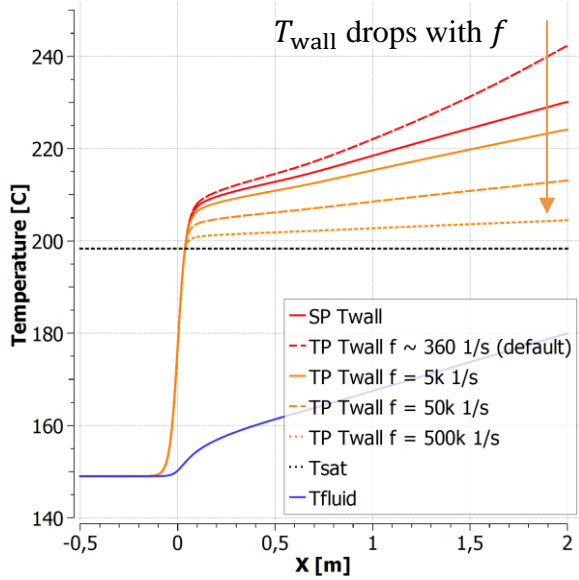
(a) Temperatures

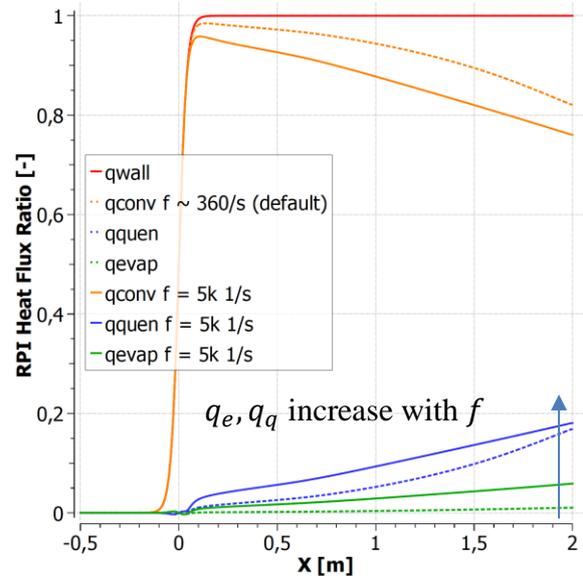
(b) Heat Fluxes

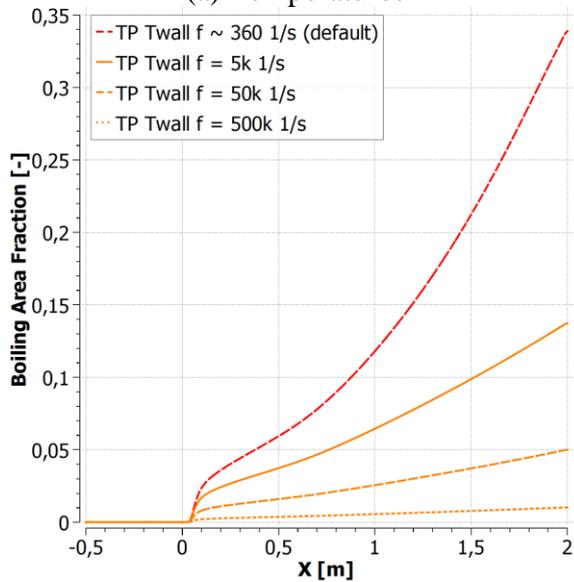
(c) Boiling Area Fraction

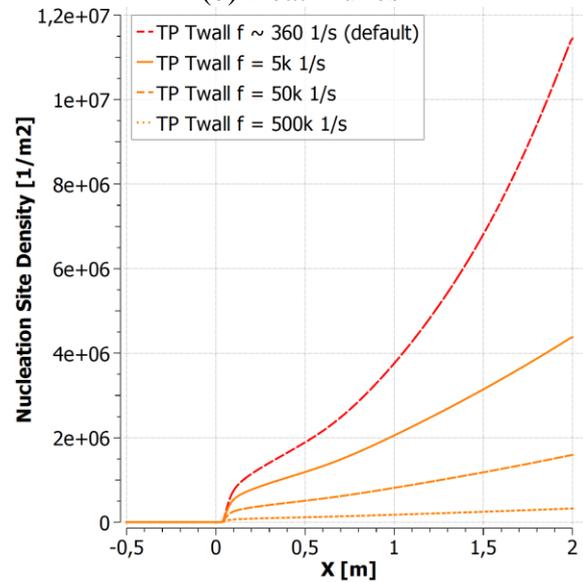
(d) Nucleation Site Density

Figure 13: Modified Bartolomei case B1 with 10 m/s and 3 MW/m$^2$ - effect of increasing bubble departure frequency $f$ on the wall temperature (a), heat partitioning (b), boiling area fraction (c) and nucleation site density (d). Labels "qwall", "qquen", "qconv", "qevap" correspond to total wall heat flux, quenching, convective, and evaporation heat flux, respectively.

### 5.3 Boiling parameters in the state-of-the art MIT heat partitioning model

Among the recently developed state-of-the-art flow boiling models, the MIT (Massachusetts Institute of Technology) model of Kommajosyula [30] seems the most promising as it addresses drawbacks related to modelling of boiling parameters in RPI heat flux partitioning model. In this section, the values of the main boiling parameters of the MIT model are compared with the results of the RPI heat flux partitioning model under divertor conditions. Mathematically, the MIT model is based on the same heat flux partitioning principle as the RPI model, but the physics background is different and hence the modelling of the main boiling parameters as described hereafter.

#### 5.3.1 Bubble departure diameter and departure frequency

In the MIT boiling model [30], the departure diameter is calculated as:

$$d_w = 18.9 \cdot 10^{-6} \, [m] \cdot \left(\frac{\rho_l - \rho_g}{\rho_g}\right)^{0.27} \cdot \mathrm{Ja}_{\sup}^{0.75} \cdot (1 + \mathrm{Ja}_{\mathrm{sub}})^{-0.3} \cdot \left(\frac{u_l}{1.0 \, [m/s]}\right)^{-0.26}. \quad (13)$$

Here, the Jakob numbers are $\mathrm{Ja}_{\sup,\mathrm{sub}} = (\rho_l \, c_{pl} \, \Delta T_{\sup,\mathrm{sub}})/\rho_g \, h_{lg}$ for the superheat ($\Delta T_{\sup} = T_{\mathrm{wall}} - T_{\mathrm{sat}}$) and subcooling ($\Delta T_{\mathrm{sub}} = T_{\mathrm{sat}} - T_l$), respectively. This model also depends on the liquid velocity scale $u_l$.

The bubble departure frequency takes into account the bubble growth and bubble waiting time

$$f = \frac{1}{t_{\mathrm{growth}} + t_{\mathrm{wait}}}, \quad (14)$$

where the waiting time is calculated as [30]

$$t_{\mathrm{wait}} = 0.0061 \, [s] \cdot \mathrm{Ja}_{\mathrm{sub}}^{0.6317} \, / \, \Delta T_{\sup} \quad (15)$$

and the bubble growth time is modelled according to Plesset and Zwick [56]

$$t_{\mathrm{growth}} = \left(\frac{d_w}{4\,K}\right)^2, \quad (16)$$

where the growth constant is

$$K = \sqrt{a_l} \cdot \mathrm{Ja}_{\sup} \cdot \left(\frac{1}{(0.804 \cdot \sqrt{\mathrm{Pr}_l})} + \left(1.55 - 0.05 \cdot \frac{\Delta T_{\mathrm{sub}}}{\Delta T_{\sup}}\right)\right), \quad (17)$$

with the liquid heat diffusivity $a_l = \lambda_l/(\rho_l c_{pl})$ and Prandtl number $\mathrm{Pr}_l = (c_{pl}\mu_l)/\lambda_l$.

In Figure 14, bubble departure diameter $d_w$ and frequency $f$ derived from MIT model are compared with the values obtained from the standard RPI model. Figure 14a shows $d_w$ values calculated by Eq. (13) at different liquid subcoolings of 50, 100, and 150 °C and wall superheat $\Delta T_{\sup}$ up to 50 °C. The wall superheat is directly related to the heat flux - a higher heat flux leads to a higher wall superheat (see Figure 9). These results are compared to the $d_w$ calculated by the Tolubinski & Kostanchuk model [37] that depends solely on liquid subcooling and does not change with the wall superheat. The flow conditions are based on the experiment of Komov et al. [11] for water at 1 MPa and mean velocity of 10 m/s. The calculations presented show that the constant bubble departure diameter of 0.1 mm used in some of the analyses in this paper is a reasonable approximation for these conditions. Experimental values of $d_w$ under similar conditions can also be found in [57].

Figure 14b compares the bubble departure frequencies calculated by the MIT [30] model using Eq. (14) with the Cole [39] model (Eq. 6) at different liquid subcooling of 50, 100, and 150

K and wall superheating. The frequency in the Cole model [39] depends on the bubble departure diameter ($f \propto 1/\sqrt{d_w}$), but not on the wall superheat. The MIT model depends on both, diameter ($f \propto 1/d_w^2$) and wall superheat ($f \propto \Delta T_{\text{sup}} + \cdots$). Both models predict a similar range for the departure frequency and departure diameters at a given liquid subcooling.

Figure 14c shows the dependency of departure diameter on the liquid velocity in the MIT model, see Eq. (13), at constant subcooling of 150 °C. Although velocity has a significant impact on the bubble departure size, this is not directly reflected in the departure frequencies shown in the following Figure 14d for different combinations of liquid velocities (10 and 1 m/s) and subcooling (5, 50 and 150 °C). As shown in Figure 15, the waiting time, which does not explicitly depend on the velocity, see Eq. (15), is much larger than the growth time in these conditions, thus the frequency is dominated by significantly longer waiting time. Nevertheless, the liquid velocity has an implicit impact on the departure dimeter and frequency, as it affects the heat transfer and hence the wall superheat and liquid subcooling at a given heat flux.

For subcooling between 50 and 150 °C, which are considered representative cases, the calculated frequency is about five times lower than the constant frequency value of 5000 Hz used to improve the performance of RPI model. However, as shown in Figure 14d, such values can be obtained with very low liquid subcooling (5 °C), which may be representative of a local liquid subcooling close to the heated wall. Not to stray too far from the discussion here, this value of $f$=5000 Hz is also used later in section 5.3.3 in the analysis of the quenching heat flux model.

To the best of our knowledge, there is no experimental data on the bubble departure frequency at divertor conditions (10 MW/m$^2$, 10 m/s, 1 MPa, 50 K superheat, 150 K subcooling). Recently, Kilavuz [58] presented a review of departure frequency models at similar operating conditions, showing values from 100 Hz to several 10 kHz for wall superheat of 40K and small bubble departure diameters $d_w$ (<0.5 mm). Recent measurements of bubble departure frequency by Kossolapov [57], performed at 1.05 MPa and mass fluxes of 500, 1000, and 2000 kg/(m$^2$s) indicate lower mean frequencies with values from ~100 to ~130 Hz at about 25K wall superheat. The distributions of frequencies range up to several kHz). But these measurements were still performed at lower flow velocities and heat fluxes than those encountered in fusion divertors (2000 vs 10000 kg/(m$^2$s), 6.15 MW/m$^2$ vs 10 MW/m$^2$).

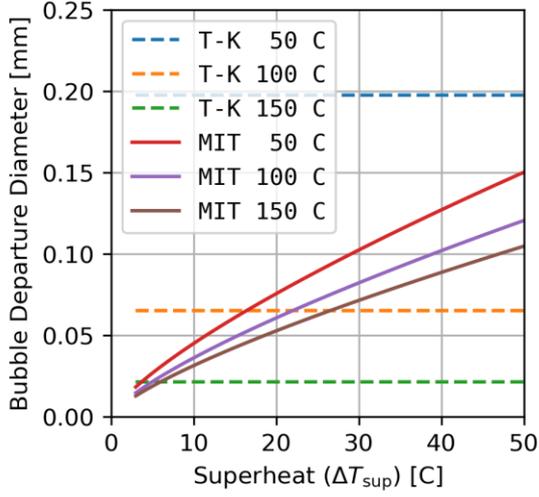
(a)

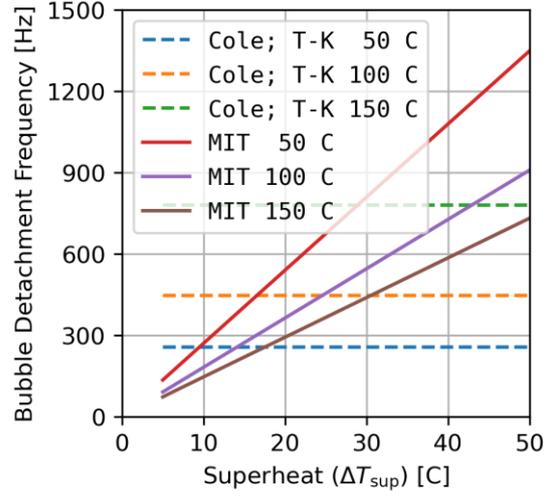
(b)

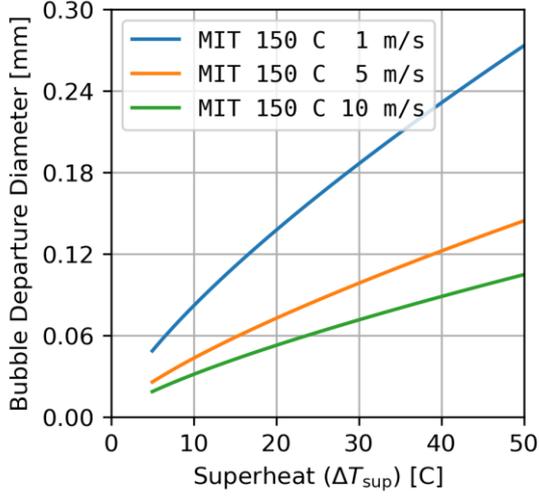
(c)

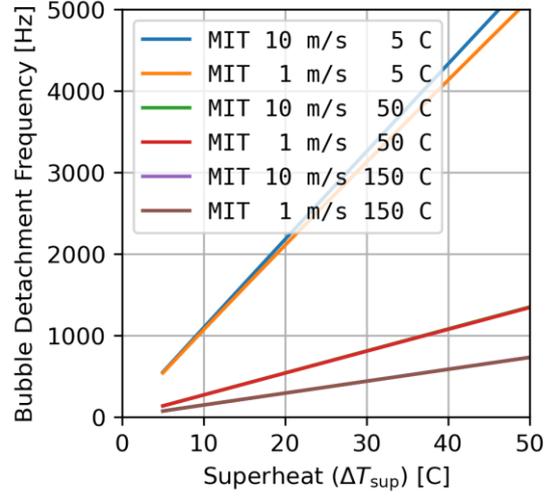
(d)

Figure 14: Comparison of models for the bubble departure diameter $d_w$ (a) and its departure frequency $f$ (b) at varying wall superheat ($T_{\text{sup}}$), calculated at different liquid subcooling of 50, 100, and 150 C. Dependency on liquid velocity in MIT model is shown in (c) at subcooling of 150 °C and velocities of 1, 5 and 10 m/s, and (d) for velocities 1 and 10 m/s and subcooling of 5, 50 and 150 °C. Models are labeled as: "T-K" - Tolubinski-Kostanchuk [37], "MIT" [30], "Cole" [39]. The flow conditions apply to water at divertor-like experiments [11] (see Table 1).

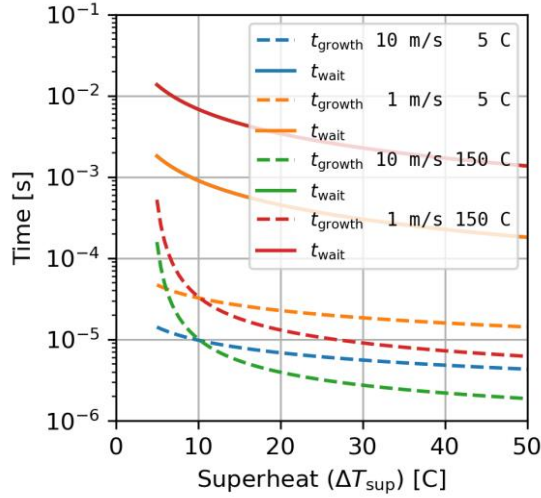

Figure 15: Bubble growth and wait times, respectively from Eq. (15) and (16) in the MIT model at different liquid velocity (10 and 1 m/s) and subcooling (5 °C and 150 °C).

### 5.3.2 Nucleation site density

To calculate the number of active nucleation sites $n_a$, MIT model [30] uses the model of Hibiki-Ishii [59] with suppression of nucleation sites due to static bubble interaction - a solution proposed by Gilman [60] for the unphysically large number nucleation sites predicted by the model. Details for this model are given in the recent literature [30], [60], [58], [61], the related equations are omitted here for brevity.

Figure 16 compares the values of $n_a$ calculated by the different models. The label "H-I" indicates the original Hibiki-Ishi model [59], the label "Mod. H-I" denotes the same model but with suppression of $n_a$ [30], the label "L-C" denotes the original Lemmert-Chawla model [38] in Eq. (8), and the label "L-C ($10^5 \cdot n_{ref}$)" denotes the Lemmert-Chawla model, where the value of $n_{ref}$ is multiplied by a factor of $10^5$ in Eq. (8). From the figure it can be seen that in the high superheat region, where the $n_a$ suppression becomes significant (in this case > 30 K), the H-I model has the same dependence on superheat (with an exponent of 1.805) as the L-C model. Note the logarithmic scale on the y-axis.

Clearly, under divertor cooling conditions (see Table 1, Komov et al. [11]), at wall superheat above 10K, the $n_a$ values calculated by the MIT model [30] (denoted as Mod. H-I) are significantly larger than in the RPI model (denoted as L-C), reaching up to five orders of magnitude ($10^5$) difference. Such high $n_a$ values would pose a challenge for the calculation of the bubble influence area fractions $A_q = \pi\, d_w^2\, n_a$ used for quenching (and similarly $A_e$ used for evaporation) in RPI model as formulated in CFX. Specifically, for values of $10^5 \cdot n_a$, the product $d_w^2 \cdot 10^5 n_a$ would immediately cause $A_q$ to be limited to the maximum value of 1 for all but the smallest (micro) bubbles $d_w < 0.01$ mm, according to values presented in Figure 14a. Arguably, the area fraction for quenching should be $A_q \leq 1$, and this is a common limit in implementations of RPI model.

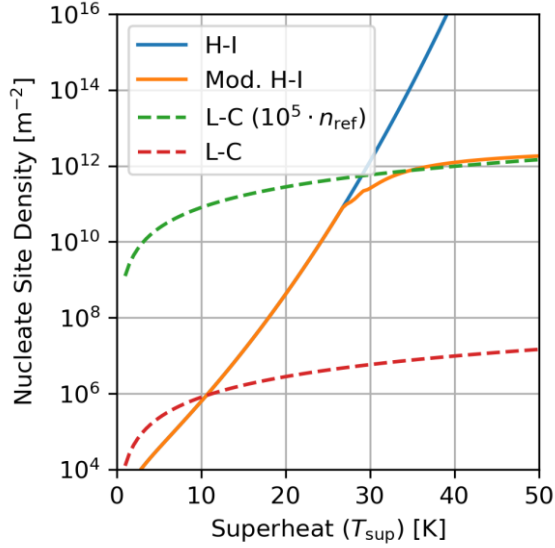

Figure 16: Comparison of calculated $n_a$ values obtained with the original model of Hibiki-Ishii (H-I), with suppression of nucleation sites $n_a$ (Mod. H-I) used in MIT model [30], with the Lemmert-Chawla model [38] (L-C) commonly used in RPI model and with and additional example using $10^5$ times higher value for the $n_{ref}$ in Eq. (8) (L-C ($10^5 \cdot n_{ref}$)). Flow conditions apply to divertor-like experiments [11] in Table 1.

### 5.3.3 Quenching heat flux model

The comparison between the MIT model [30] and classical RPI model [27] continues with the most significant difference between them: the modelling of quenching in the RPI model, which is replaced by the bubble sliding model in MIT formulation. In the bubble sliding model [30], the bubble influence area fraction in the RPI model is replaced by the sliding area fraction defined as

$$A_{sl} = \min(1, S_{sl}\, n_a\, t^* f). \tag{18}$$

Here, $S_{sl} = 1.1 \cdot d_w/\sqrt{n_a}$ is the bubble sliding surface area in m², and $t^* = 1/(\pi a_l) \cdot (\lambda_l/h_{cl})^2$ is the thermal boundary layer re-establishment time [30]. Following the example presented in [57], both quenching models can be rewritten in a similar form for comparison:

$$q_{q,\ \text{RPI [27]}} = \pi d_w^2\, n_a \quad t_w f \quad \frac{2}{\sqrt{\pi\, t_w}} \cdot \frac{\lambda_l}{\sqrt{a_l}} \quad (T_w - T_l) \tag{19}$$

$$q_{q,\ \text{MIT [39]}} = 1.1 \cdot d_w \sqrt{n_a} \quad t^* f \quad \frac{2}{\sqrt{\pi\, t^*}} \cdot \frac{\lambda_l}{\sqrt{a_l}} \quad (T_w - T_l) \tag{20}$$

for brevity omitting the constraints on minimum bubble influence area fraction and assuming the default value of $F_2 = 2$ in RPI quenching heat flux model.

Note that the product $S_{sl} n_a = 1.1 \cdot d_w \sqrt{n_a}$ also represents a dimensionless fraction of bubble influence area (where sliding occurs) that stems from spatial averaging (similar to $\pi d_w^2\, n_a$ in RPI). The other product pair $t_w f$ and $t^* f$ represents a time fraction, assuming $1/f$ is the time between the departing bubbles. The commonly used value for the waiting time in the RPI model is the default constant value of the waiting time fraction $t_w f = 0.8$. Unlike in RPI, the time fraction in the MIT model [30] varies dynamically, limited only by the maximum fraction of heated surface coverage ($A_{sl} = 1$), as shown in Eq. (18).

The third part of the form $h'_{t_w} = 2\lambda_l/\sqrt{\pi t_w a_l}$ for the RPI model and $h'_{t^*} = 2\lambda_l/\sqrt{\pi\, t^* a_l}$ for the MIT model represents the quenching HTC with the units of (W/m²K) and can be viewed as an upper limit for quenching at a given temperature difference. Hence, the two models differ only in that the bubble waiting time $t_w$ in RPI model is replaced by the thermal boundary layer re-establishment time $t^*$ in the MIT model.

As defined by Del Valle and Kenning [45] and shown in Eq. (7), the quenching $h_q$ includes the time fraction and is defined as $h_q = t_w f \cdot h'_{t_w}$ using the above HTC part. Assuming a constant $t_w f = 0.8$, the quenching in RPI depends on the departure frequency: $h_q \propto 1/\sqrt{t_w} \propto \sqrt{f}$.

For the MIT sliding model [30]**Error! Reference source not found.**, their presented example of CFD implementation derives the sliding HTC as:

$$h_{q,\text{ MIT [39]}} = \frac{2}{\sqrt{\pi\, t^*}} \cdot \frac{\lambda_l}{\sqrt{a_l}} = 2\, h_{cl}, \tag{20}$$

which follows from the definition of $t^*$. Compared to RPI, this definition excludes the time fraction $t^* f$ from the HTC part, and groups it with sliding area fraction, as shown in Eq. (18). This is an important difference with RPI, since the bubble influence area fraction $A_q$ and sliding fraction $A_{sl}$ also determine the area fraction used for the convective heat flux, as $(1 - A_q)$ or $(1 - A_{sl})$. This feature eliminates the issue with quenching modelling in RPI, presented in section 5.1, namely a too low quenching HTC in Eq. (7). At a given boiling influence area fraction (or sliding fraction), the convective HTC is essentially replaced with twice as effective quenching HTC in MIT model.

To explore the dependence of the models on the bubble departure frequency and to facilitate comparison between the two models, the following ratio is derived from Eqs. (7) and (20):

$$\frac{h_q}{2 h_{cl}} = \frac{t_w f / \sqrt{t_w}}{1/\sqrt{t^*}} = \sqrt{(t_w f)(t^* f)}, \tag{21}$$

taking into account that the part with physical properties $2\lambda_l/\sqrt{\pi a_l}$ is the same in both models and cancels out; the remainder represents the different time fractions in the models. Since the product $t_w f$ is assumed to be constant, the ratio depends solely on $t^*$, that is based on convective $h_{cl}$, and departure frequency $f$. Figure 17 thus shows the square root dependency of the ratio in Eq. (21) with respect to frequency at three different $t^*$ of 0.5, 0.2, and 0.13 ms, which correspond to convective $h_{cl}$ of approximately 20, 40, and 80 kW/m²K, respectively, typical for the divertor cooling conditions (experiments of Komov et al [11]). The solid lines show the calculated values at $t_w f = 0.8$ and the dashed lines at $t_w f = 1$. The filled dots indicate the frequencies corresponding to the $t^* f = 1$.

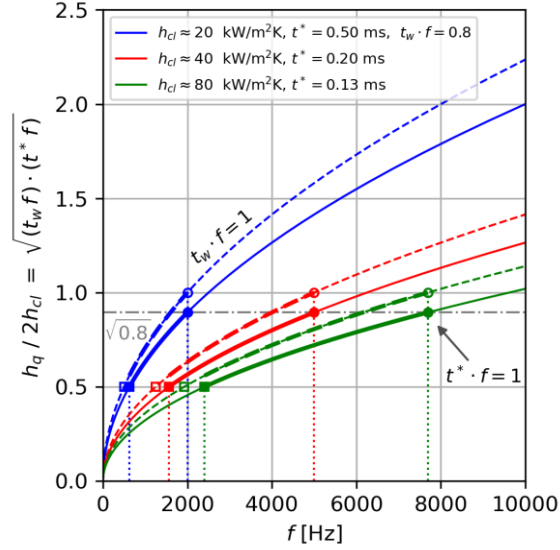

Figure 17: The ratio between quenching $h_q$ in Eq. (7) used in the RPI model and $2h_{cl}$ in Eq. (20) in the MIT model, at three different thermal boundary layer re-establishment times $t^*$ 0.5, 0.2, 0.13 ms, corresponding approximately to convective HTC values of 20, 40, and 80 kW/m²K, respectively. The dashed lines show the ratios assuming quenching time fraction $t_w f$ of 1 instead of 0.8. The squares and dots indicate lower and upper frequency limits, respectively. Note at $t^* f = 1$, the ratio $\sqrt{t_w f}$ is equal to $\sqrt{0.8}$.

Frequencies with $t^* f > 1$ indicate a bubble departure timescale that is shorter than the thermal boundary layer re-establishment time, which is probably unrealistic and therefore at $t_w$ equal to $t^*$ the product $t^* f = 1$ represents a reasonable upper limit for the frequency, $f_{\max} = 1/t^*$, or 5000 Hz at 0.2 ms. At the upper frequency limit ($t^* f = 1$), the value of the ratio in Eq. (21) is equal to $\sqrt{t_w f}$, i.e. either $\sqrt{0.8}$ or 1 for this case, and the quenching HTC is $h_q = 2\, h_{cl}\sqrt{t_w f}$, or $h_q = 2\, h_{cl}\sqrt{t_w/t^*}$. By considering the same values for $t_w$ and $t^*$, the upper bound for $h_q$ is then equal to $h_q = 2\, h_{cl}$.

To obtain the lower bound on the bubble frequency and avoid the issues mentioned in Section 5.1, the quenching has to be at least as effective as in the case of single-phase convection, i.e. $h_q$ shall be at least equal to $h_{cl}$. From Eq. (21) this gives a lower frequency limit of $f_{\min} = 1/4 \cdot 1/(t_w f) \cdot 1/t^*$. With $t_w f = 1$ the lower limit is 1/4 of the maximum frequency $f_{\max}$, and for the default case (with $t_w f = 0.8$) it is $1/(4 \cdot 0.8)$, or approximately 0.31 of the upper frequency limit. For the reference case with 40 kW/m²K and 0.2 ms, the lower frequency limit is either 5000/4 = 1250 Hz at or approximately 1500 Hz at $t_w f = 0.8$.

In summary, the presented analysis of the quenching model gives an approximate range of bubble departure frequencies between 1250 and 5000 Hz for the divertor-like conditions, corresponding to the value for the convective HTC of 40 kW/m²K in these conditions. By introducing these higher frequencies to the RPI model, the quenching heat flux in RPI model mimics the higher HTC of the MIT sliding model [30]. The simulations with the modified RPI model using the departure frequency of 5000 Hz are presented in the next section.

### 5.4 Simulation of the divertor conditions with the modified RPI heat partitioning model

Following the preceding sensitivity analysis, the RPI model has been modified by using an increased bubble departure frequency $f$, which alleviates the issue of unphysical wall temperature prediction and may considerably improve the simulation of boiling under divertor conditions.

In the "modified" RPI heat partitioning model, the bubble departure frequency is set to a constant value of $f = 5000$ Hz and the bubble departure diameter $d_w$ is calculated using the Unal model [62] with an upper limit of 0.1 mm. The modified RPI heat partitioning model was used to simulate the W7-X divertor cooling channel and the experiments of Komov et al. [11]. The simulations with the modified and base RPI model are compared with the single-phase results and experiments in Figure 18. The simulation "TP base" denotes the results obtained with the base RPI heat partitioning model described in section 2.2, while the simulation "TP modified" denotes the results with the modified RPI model.

Simulations of W7-X divertor cooling channel are shown in Figure 18 (a). As can be seen, the modified model with the departure frequency of 5000 Hz predicts a lower wall temperature than the single-phase simulation, thus providing a physically plausible result. As already discussed in the previous section, the simulation with a frequency of 1000 Hz (blue dashed curve) still overpredicts the SP wall temperature. The simulation with the base model where the frequency is around 184 Hz is clearly incorrect.

Figure 16 (b) provides the simulations of Komov experiments [11]. Both TP simulations and the SP simulation coincide and agree with the experimental results as long as the wall temperature remains below the saturation temperature, i.e. when only single-phase convection is present. At higher heat fluxes, in the boiling regime, the "TP base" simulation predicts a higher wall temperature than the SP simulation, and both are much higher than the experimental values. Similar to the divertor case, the "TP base" results are treated as unphysical. On the contrary, the simulations with the "TP modified" model show a good agreement with the experimental data and correctly predict the decline of the wall temperature slope in the case of boiling. A relatively simple modification of the RPI model can therefore give physically meaningful results.

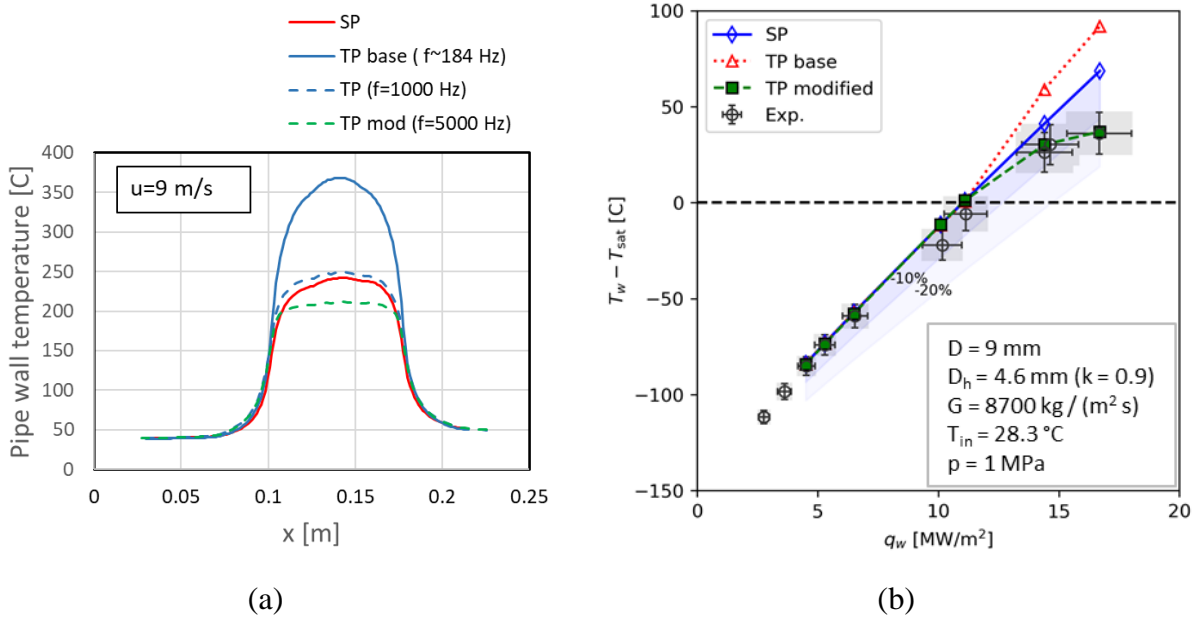

(a)  (b)

Figure 18: Simulations of the W7-X target element (a) and experiments of Komov et al. [11] (b) using the modified and base RPI model; Single-phase (SP) result is provided for reference.

However, developing a mechanistic heat partitioning model that also appropriately reflects the physical mechanisms at high heat flux and high flow velocity conditions is still needed. This is not possible without validation on CFD-grade experiments. Dedicated flow boiling visualization experiments are essential to understand the main boiling mechanisms and gather key information on boiling parameters for the CFD closure models.

## 6   Conclusions

The ability of existing CFD models to predict flow boiling at high heat fluxes and high flow velocities was analyzed within the ANSYS CFX framework. It was shown that existing boiling models used in the vast majority of CFD codes produce unphysical results (overprediction of wall temperature) in cooling channels of the divertor target elements at heat fluxes around 10 MW/m$^2$ and high coolant flow velocities of around 10 m/s. The CFD simulations of flow boiling were validated on the available experimental data in uniformly and top-heated flow channels. It has been shown that the boiling model correctly predicts the void fraction and wall temperature at low velocities below 3m/s, but fails to provide physical meaningful results at higher flow velocities. The main reason for the erroneous predictions was found to be the RPI heat flux partitioning model. The model was analysed in detail, focusing on the effects of different heat flux contributions and boiling parameters, such as bubble departure diameter $d_w$, bubble departure frequency $f$ and nucleation site density $n_a$.

The main findings of the analyses can be summarized as: the overall heat transfer coefficient between the wall and the boiling fluid is lower (unphysical) in the boiling simulation than in the conservative single-phase simulation, resulting in a higher wall temperature in the boiling (two-phase) case. The modelling of quenching heat flux was recognized to be the main source of misprediction as it is rather insensitive to the change in flow velocity. This can be attributed to the models for the bubble departure frequency $f$ and the bubble waiting time $t_w$, which are originally derived from the pool boiling and do not properly account for the effects of higher flow velocity. To address this issue, the RPI heat flux partitioning model was compared to the recent MIT model [30], which is based on similar heat flux distribution but considers bubble sliding. Based on thorough analysis and comparison with the MIT model, a proposed modification of the RPI model introduces a higher bubble departure frequency, which adequately mimics the higher heat transfer under divertor-like conditions.

The study identifies the need to re-evaluate some modelling assumptions, which should be based on CFD-grade flow boiling experiments, and to develop an improved heat partitioning model. Due to the lack of relevant data, there is a clear need for the development of dedicated experimental setups to study the fundamental boiling mechanisms and to obtain the main modelling parameters.

**Acknowledgements**

This work has been carried out within the framework of the EUROfusion Consortium, funded by the European Union via the Euratom Research and Training Programme (Grant Agreement No. 101052200 - EUROfusion). Views and opinions expressed are however those of the author(s) only and do not necessarily reflect those of the European Union or the European Commission. Neither the European Union nor the European Commission can be held responsible for them.

The financial support from the Slovenian Research and Innovation Agency grants P2-0405 and P2-0026 is gratefully acknowledged.